\documentclass{article}

\usepackage{arxiv}

\usepackage[utf8]{inputenc} 
\usepackage[T1]{fontenc}    
\usepackage{hyperref}       
\usepackage{url}            
\usepackage{booktabs}       
\usepackage{amsfonts}       
\usepackage{amssymb}
\usepackage{nicefrac}       
\usepackage{microtype}      
\usepackage{graphicx}
\usepackage{subcaption}
\usepackage{natbib}
\usepackage{amsmath}
\usepackage{doi}
\usepackage[title]{appendix} 
\usepackage{tikz}
\setcitestyle{square,numbers,comma}

\usetikzlibrary{shapes.geometric, arrows, arrows.meta, positioning, fit, calc, backgrounds, shadows, decorations.pathreplacing}

\title{Profiling systematic uncertainties in Simulation-Based Inference with Factorizable Normalizing Flows}

\author{ \href{https://orcid.org/0000-0001-8587-8266}{\includegraphics[scale=0.06]{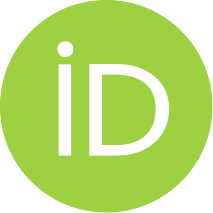}\hspace{1mm}Davide Valsecchi} \\
	D-PHYS Department \\
	ETH Zurich\\
	Zurich, Switzerland \\
	\texttt{dvalsecchi@ethz.ch} \\
	\And
	\href{https://orcid.org/0000-0001-9830-0412}{\includegraphics[scale=0.06]{orcid.pdf}\hspace{1mm}Mauro Doneg\`a} \\
	D-PHYS Department \\
	ETH Zurich\\
	Zurich, Switzerland \\
	\texttt{mdonega@ethz.ch} \\
	\AND
	\href{https://orcid.org/0000-0001-8038-1613}{\includegraphics[scale=0.06]{orcid.pdf}\hspace{1mm}Rainer Wallny} \\
	D-PHYS Department \\
	ETH Zurich\\
	Zurich, Switzerland \\
	\texttt{rwallny@ethz.ch} \\
}

\renewcommand{\headeright}{Preprint}
\renewcommand{\undertitle}{Preprint}
\renewcommand{\shorttitle}{Profiling systematic uncertainties in SBI with Factorizable Normalizing Flows}

\hypersetup{
pdftitle={Profiling systematic uncertainties in Simulation-Based Inference with Factorizable Normalizing Flows},
pdfsubject={},
pdfauthor={Davide Valsecchi, Mauro Doneg\`a, Rainer Wallny},
pdfkeywords={Simulation-Based Inference, Normalizing Flows, Amortized Inference, Profile Likelihood, Systematic Uncertainties, Unbinned Analysis, High Energy Physics},
}

\begin{document}
\maketitle

\begin{abstract}
Unbinned likelihood fits maximize the information extracted from experimental data, yet their application in realistic high-dimensional analyses has been fundamentally bottlenecked by the prohibitive computational cost of profiling systematic uncertainties. Furthermore, current machine learning-based inference methods typically estimate scalar parameters, discarding complex high-dimensional correlations. To address this, we propose a general Simulation-Based Inference (SBI) framework that elevates the fit target from scalar parameters to a multivariate \emph{Distribution of Interest} (DoI), a learnable, invertible transformation of the feature space. We employ Factorizable Normalizing Flows to model systematic variations as parametric deformations, preserving tractability without combinatorial explosion. Crucially, we develop an amortized training strategy that learns the conditional dependence of the DoI on nuisance parameters in a single optimization process, bypassing repetitive training during likelihood scans. To capture the finite-sample statistical variance of the neural network DoI, we introduce a Poisson-bootstrap ensemble, which we marginalize through an averaged likelihood to deliver a complete statistical-plus-systematic uncertainty budget within a single unbinned likelihood. Validated on a synthetic dataset emulating a high-energy physics measurement, our method demonstrates that rigorous, fully profiled unbinned measurements can now be extended to complete differential distributions. By turning the fit into a functional measurement, this approach offers a powerful, unifying framework for a broad range of tasks conventionally treated as distinct problems, from detector calibration and differential cross-sections to unfolding and continuous parameter estimation.
\end{abstract}



\section{Introduction}
\label{sec:intro}

Unbinned likelihood fits have long been recognized as a powerful statistical tool in High Energy Physics (HEP) data analysis, offering the potential to fully exploit the information contained in continuous distributions of observables. Unlike binned approaches, which discretize data into histograms and can suffer from information loss and binning artifacts, unbinned fits operate directly on the raw event data, enabling more precise parameter estimation and hypothesis testing. Two obstacles have nonetheless limited their use in realistic LHC analyses: the measurement is almost always reduced to a small set of scalar parameters, and the profiling of systematic uncertainties becomes computationally prohibitive in high-dimensional feature spaces.

The first of these obstacles reflects how the measurement itself is framed: most existing unbinned methods target scalar quantities, such as signal strengths or Wilson coefficients~\cite{Schofbeck:2025smeft,Barrue:2026qul,Brehmer:2018eca,ATLAS:2025nsbi,Amram:2025hisigma}, leaving the more general problem of measuring a full \emph{differential distribution}, crucial for cross-section measurements and generator tuning, largely unaddressed. The second concerns the treatment of systematic uncertainties. Traditionally, these are handled by modelling their effect on the observables as a variation around a nominal template, often constructing histograms for every ``up'' and ``down'' variation of the nuisance parameters. While effective and scalable to hundreds of parameters in binned fits with one or a few parameters of interest~\cite{Cranmer:2012sba,CMS:2024combine}, this strategy struggles in high-dimensional feature spaces where binning becomes intractable. Recent advances in Simulation-Based Inference (SBI)~\cite{Cranmer:2020wdu,Held:2025advancing} approximate likelihood ratios with neural networks and operate naturally in high dimensions, but incorporating systematic uncertainties remains a significant hurdle: many strategies~\cite{ATLAS:2025calib,Algren:2025gap,ATLAS:2025nsbi,Schofbeck:2025smeft} require either retraining models for each systematic variation or training computationally expensive conditional density estimators that must explicitly learn the effect of every nuisance parameter over its full range.

This work makes three contributions. The first is conceptual: we generalize the target of the fit from a set of scalar parameters to a full \emph{Distribution of Interest} (DoI). Rather than summarizing the data into a few numbers, the measurement becomes a learnable, invertible transformation $T_{\phi}$ of the feature space that maps a reference model onto the observed data. Since the transformation is a diffeomorphism with a tractable Jacobian, it defines a valid probability density at every step and can be optimized directly inside an unbinned likelihood. The fitted transformation is itself the result, and many measurements that are conventionally treated as distinct problems become instances of a single procedure: the measurement of an invertible map between two feature spaces. The procedure delivers a continuous detector calibration when the two spaces are simulation and data, a differential cross-section when the learned map is read off as a function of the observables, unfolding when the two spaces are the reconstructed detector-level and the true particle-level distributions, and ordinary parameter estimation when the target reduces to a single continuous quantity such as a mass or a coupling. In this sense the DoI turns an unbinned fit into a functional measurement that preserves the high-dimensional correlations a scalar parametrization or a binned histogram would discard, and offers a unifying framework for a large class of the measurements performed at the LHC.

The second contribution makes such a measurement realistic: systematic uncertainties must be profiled, and profiling a functional target is qualitatively different from the scalar case. In a standard fit one scans the parameter of interest and, at each value, maximizes the likelihood over the nuisance parameters; a distribution cannot be scanned in this way. We instead exploit the symmetry of the profiling operation and swap the roles of the parameters, learning the best-fit transformation as a function of the nuisances and propagating their data-constrained variation into the measurement. Establishing that this inverted construction is a genuine profiling of systematic uncertainties, and making it computationally tractable, is a central methodological contribution of this work; the precise formulation is given in Section~\ref{sec:systematics}.

Making this profiling tractable rests on two ingredients. To model systematic effects we adopt the \emph{Factorizable Normalizing Flows} (FNF) of Ref.~\cite{Valsecchi:2026fnf}, which structurally decompose the probability density into a nominal component and learnable, invertible deformations, capturing continuous high-dimensional systematic variations with a cost that scales linearly, rather than exponentially, with the number of nuisance parameters. On top of this we introduce an \textbf{amortized optimization strategy} that learns the response of the likelihood to the nuisance parameters across their whole range in a single training, replacing the repeated fits of a conventional profiling scan with a one-time upfront cost.

The third contribution addresses the statistical uncertainty of the measurement itself. Because the Distribution of Interest is a neural network fitted to a finite dataset, the learned transformation would fluctuate if the measurement were repeated on independent data; since the transformation is itself the result, this finite-sample variability is a genuine statistical uncertainty that must be propagated alongside the systematic one. We capture it with a \textbf{Poisson-bootstrap ensemble} of the fit, whose spread samples the statistical fluctuation of the learned map, and combine it with the systematic profiling through an \textbf{ensemble-averaged (bagged) likelihood}. The measurement therefore delivers a complete statistical-plus-systematic uncertainty budget for a functional, neural-network target, obtained within a single unbinned likelihood rather than assembled by hand after the fit.

This paper is structured as follows: Section~\ref{sec:formulation} details the mathematical formulation, modelling probability densities with Normalizing Flows and defining the DoI framework. Section~\ref{sec:systematics} recalls the Factorizable Normalizing Flows of Ref.~\cite{Valsecchi:2026fnf} for systematic handling, introduces the amortized profiling strategy, and develops the Poisson-bootstrap propagation of the statistical uncertainty of the learned transformation. In Section~\ref{sec:experiments}, we validate the method on a synthetic physics dataset with multiple systematic sources. Finally, Section~\ref{sec:discussion} discusses the outlook and potential applications of this framework in real-world HEP analyses.

\begin{figure}[!htb]
    \centering
    \resizebox{\textwidth}{!}{%
    \begin{tikzpicture}[font=\sffamily]

    \tikzset{
        databox/.style={
            cylinder, shape border rotate=90, draw=black!80, fill=yellow!10, very thick,
            aspect=0.25, minimum height=1.5cm, minimum width=2.0cm, align=center, font=\small,
            drop shadow={opacity=0.15}
        },
        fixed/.style={
            rectangle, rounded corners, draw=black!50, fill=gray!10, very thick, dashed,
            minimum width=4cm, minimum height=2.0cm, align=center, font=\small,
            drop shadow={opacity=0.15}
        },
        flow_meas/.style={
            rectangle, rounded corners, draw=black!80, fill=blue!10, very thick,
            minimum width=4cm, minimum height=2.0cm, align=center, font=\small,
            drop shadow={opacity=0.15}
        },
        flow_ref/.style={flow_meas},
        systematic/.style={
            rectangle, rounded corners, draw=black!80, fill=green!15, very thick,
            minimum width=4cm, minimum height=2.0cm, align=center, font=\small,
            drop shadow={opacity=0.15}
        },
        nuisance/.style={
            diamond, draw=black!80, fill=orange!50!red!35!white, aspect=2, inner sep=2pt, thick, font=\large,
            drop shadow={opacity=0.15}
        },
        endpoint/.style={
            circle, draw=black!80, fill=orange!20, thick, align=center, font=\large\bfseries,
            drop shadow={opacity=0.15}
        },
        flow/.style={->, >=stealth, very thick, black!70},
        cond/.style={->, >=stealth, thick, dashed, orange!50!red}
    }

    
    \node[databox] (data) at (0,0) {\textbf{Observed}\\\textbf{Data} $x$};
    
    \node[systematic] (sys_doi) at (6,0) {
        \textbf{DoI Systematics}\\
        $T_{\psi}(\cdot \mid \nu)$\\
        {\scriptsize\itshape\color{black!55}Uncertainty on Map}
    };

    \node[fixed, drop shadow={opacity=0}] at (12.34,0.34) {};
    \node[fixed, drop shadow={opacity=0}] at (12.17,0.17) {};
    \node[fixed] (nom_doi) at (12,0) {
        \textbf{Ensemble of}\\\textbf{Nominal DoI}\\
        $\{T_{\phi}^{(b)}\}$\\
        {\scriptsize\itshape\color{black!55}Physics Map}
    };
    
    \node[flow_ref, minimum width=4cm] (sys_input) at (12,-4.5) {
        \textbf{Input Systematics}\\
        $T_{\chi}(\cdot \mid \nu)$\\
        {\scriptsize\itshape\color{black!55}Uncertainty on Ref}
    };

    \node[fixed, dashed] (density) at (6,-4.5) {
        \textbf{Nominal Density}\\
        $p_{\text{nom}}(z)$\\
        {\scriptsize\itshape\color{black!55}Fixed Template}
    };

    \node[endpoint, label={[font=\scriptsize\itshape, text=black!55]below:Likelihood}] (nll) at (0,-4.5) {$\mathcal{L}$};

    
    \node[nuisance] (nu) at (9,-2.25) {$\nu$};
    \node[font=\scriptsize, text=black!55, left=0.2cm of nu] {Nuisance parameters};

    
    \draw[flow] (data) -- (sys_doi);
    \draw[flow] (sys_doi) -- (nom_doi) node[midway, above=0.1cm, font=\scriptsize, text=black!55, align=center] {Corrected\\Data};
    \draw[flow] (nom_doi.south) -- (sys_input.north) node[midway, right=0.1cm, font=\scriptsize, text=black!55, align=left] {Reference\\Space};
    \draw[flow] (sys_input) -- (density) node[midway, below=0.1cm, font=\scriptsize, text=black!55, align=center] {Nominal\\Space};
    \draw[flow] (density) -- (nll);

    \draw[cond] (nu) -- (sys_doi.south east);
    
    \draw[cond] (nu) -- (sys_input.north west);
    
    \begin{scope}[on background layer]
        \node[fit=(sys_doi)(nom_doi), draw=gray!30, fill=white, very thick, rounded corners, inner sep=0.5cm, drop shadow={opacity=0.1}, label={[text=black!70, font=\footnotesize\scshape]above:Distribution of Interest (Measurement)}] {};
        
        \node[fit=(sys_input)(density), draw=gray!30, fill=white, very thick, rounded corners, inner sep=0.5cm, drop shadow={opacity=0.1}, label={[text=black!70, font=\footnotesize\scshape]below:Reference Model (Background/Signal)}] {};
    \end{scope}

    \end{tikzpicture}
    }%
    \caption{Conceptual overview of the proposed framework. The observed data $x$ are carried back through the learnable transformations to the frozen nominal density $p_{\text{nom}}$, where the likelihood $\mathcal{L}$ is evaluated. The measurement is the invertible \textbf{Distribution of Interest} (DoI) $T_{\phi}$ that morphs the data onto this nominal reference. Systematic uncertainties, conditioned on the nuisances $\nu$, enter through two flows: an \textbf{input systematic flow} $T_{\chi}$ that deforms the reference model (calibrated on simulation), and a \textbf{DoI systematic flow} $T_{\psi}$ that deforms the measurement (fixed by profiling). The nominal DoI is a \textbf{Poisson-bootstrap ensemble} $\{T_{\phi}^{(b)}\}$ whose member spread propagates the statistical uncertainty.}
    \label{fig:concept_graph}
\end{figure}

\section{Formulation}
\label{sec:formulation}

We model the observed-data density with Normalizing Flows and then promote the target of the fit from scalar parameters to an invertible transformation of the feature space.

\subsection{Unbinned likelihood and density estimation}
\label{sec:methodology}

We assume the data consist of independent events drawn from a set of physics processes, or ``flavours'', indexed by $f \in \{1, \dots, F\}$, and express the total density as a mixture model:
\begin{equation}\label{eq:mixture_model}
    p(x | \nu) = \sum_{f=1}^F \eta_f(\nu) \, p_f(x | \nu)
\end{equation}
where the $\eta_f$ are the per-flavour yields, which may depend on the nuisance parameters $\nu$, and the per-flavour densities $p_f(x|\nu)$ may themselves depend on $\nu$.

We model each $p_f(x|\nu)$ with a Normalizing Flow (NF)~\cite{Kobyzev:2020nf}, which allows the exact evaluation of the density by learning an invertible map to a simple base distribution. These reference densities are pre-trained on simulated events and kept fixed during inference, acting as the static components of the mixture.

We split the feature space into kinematic variables $x$ (e.g. $p_T$, $\eta$) and discriminative variables $y$ (e.g. classifier outputs, reconstructed masses), and factorize each component density as
\begin{equation}
  p_f(y, x | \nu) = p_f(y | x, \nu) \, p_f(x | \nu)
\end{equation}
using a conditional flow for $p_f(y | x, \nu)$ and a separate flow for the kinematic density $p_f(x | \nu)$.

\paragraph{Likelihood}
The fit maximizes the \textbf{joint likelihood} over the full feature space $(x, y)$, with negative log-likelihood
\begin{equation}
  -\ln \mathcal{L}_{\text{joint}} = - \sum_{i=1}^N \ln \left( \sum_{f} \eta_f \, p_f(y_i | x_i) p_f(x_i) \right)
\end{equation}
To constrain the total yield we adopt the extended formalism, which multiplies the per-event density by the Poisson probability of observing $N$ events when $\mu = \sum_f \eta_f$ are expected,
\begin{equation}
  \mathcal{L}_{\text{ext}} = \frac{\mu^N e^{-\mu}}{N!} \prod_{i=1}^N p(x_i, y_i)
\end{equation}
so that the negative log-likelihood reads
\begin{equation}
  -\ln \mathcal{L}_{\text{ext}} = -\sum_{i=1}^N \ln \left( \sum_{f} \eta_f \, p_f(y_i | x_i) p_f(x_i) \right) + \mu - N \ln \mu + \ln N!
\end{equation}
where the constant $\ln N!$ is dropped during optimization. This fits shape and normalization simultaneously, as is standard in HEP analyses.

\paragraph{Conditional likelihood}
Alternatively, the fit can use only the information contained in $y$ conditional on $x$,
\begin{equation}
  -\ln \mathcal{L}_{\text{cond}} = - \sum_{i=1}^N \ln p(y_i | x_i) = - \sum_{i=1}^N \ln \left( \sum_{f} w_{f}(x_i) \, p_f(y_i | x_i) \right)
\end{equation}
where the posterior weights
\begin{equation}
  w_f(x) = \frac{\eta_f p_f(x)}{\sum_{k} \eta_k p_k(x)} = P(\text{flavour}=f | x)
\end{equation}
decouple the modeling of $p(x)$ from the fit and can be estimated with a flavour classifier~\cite{Cranmer:2020wdu,Andreassen_2020}. A hybrid treatment is also possible, modeling part of the kinematics explicitly to constrain the systematic variations in $\nu$ while treating the rest conditionally.

\subsection{\emph{Distributions of Interest} as fit targets}

While the mixture model provides a baseline for standard parameter estimation, where $\eta_f$ and $\nu$ are optimized, we instead treat an \textbf{invertible transformation} that morphs the reference model to match the data as the \textbf{measurement target itself}. Conventional unbinned fits optimize a small set of scalar parameters, such as signal strengths or masses, given a fixed density model~\cite{Cranmer:2020wdu,ATLAS:2025nsbi}. We extend the optimization domain to a \textbf{family of invertible transformations} $T_\phi$, parameterized by neural networks, which we call \textbf{Distributions of Interest} (DoI): they generalize the measurement of a single parameter to that of a full distribution, defined implicitly by $T_\phi$.

We model the observed data by composing a reference density $p_{\text{ref}}(u)$, built from the mixture model of Eq.~\ref{eq:mixture_model}, with a learnable diffeomorphism $u = T_\phi(x)$, so that the fit also learns the morphing of the observable space. The likelihood is then a functional of the transformation:
\begin{equation}
    \mathcal{L}(T_\phi) = \prod_{i=1}^N p_{\text{ref}}(T_\phi(x_i)) \left| \det \nabla_{x} T_\phi(x_i) \right|
\end{equation}
The transformation must be invertible with a tractable Jacobian determinant, so that the morphed density enters the likelihood directly; its invertibility also lets us map the reference back to the target after the fit, for example to correct simulation to data. The transformation is conditioned on the event flavour, allowing process-specific corrections; we leave this dependence implicit and write $T_\phi$ throughout. It is realized as an invertible neural network whose concrete architecture is given in Section~\ref{sec:experiments}.

In our application the transformation acts on the feature space $y$, conditionally on the kinematic variables $x$ and the nuisances $\nu$, giving the per-flavour density
\begin{equation}
p_f(y, x | \nu) = p_f(T_{\phi}(y | x, \nu))  \,  p_f(x | \nu) \, \cdot \left| \det \nabla_{y   } T_{\phi}(y | x, \nu) \right|
\end{equation}
It is set up to bring the \emph{data space} $y'$ to the \emph{simulation space} $y$, with the inverse matching the distortion map applied to the data,
\begin{equation}
  T_{\phi}(y' | x, \nu) = y \quad \Rightarrow \quad y' = F(y | x, c)
\end{equation}
The same construction applies to the full feature space $(x, y)$ or to the kinematic space $x$ alone, depending on the analysis.

\paragraph{Maximum likelihood fit}
The full extended likelihood is a function of the DoI $T_{\phi}$ and the nuisances $\nu$:
\begin{equation}
  \mathcal{L}_{\text{ext}}(y, x \mid \nu, T_{\phi}) = \frac{\mu^N e^{-\mu}}{N!} \prod_{i=1}^N \left( \sum_{f} \eta_f(\nu) \, p_f(T_{\phi}(y_i | x_i, \nu))  \,  p_f(x_i | \nu) \, \cdot \left| \det \nabla_{y   } T_{\phi}(y_i | x_i, \nu) \right|  \right)
\end{equation}
The input densities $p_f(x | \nu)$ and $p_f(y | x, \nu)$ are pre-trained on simulation and frozen, while $T_{\phi}$ and $\nu$ are optimized jointly in the fit,
\begin{equation}
  \hat{T}_{\phi}, \hat{\nu} = \arg \max_{\phi, \nu} \mathcal{L}_{\text{ext}}(y, x \mid \nu, T_{\phi})
\end{equation}
We discuss the training dynamics and the handling of nuisance parameters in Section~\ref{sec:training_dynamics}.

\section{Handling of systematic uncertainties}
\label{sec:systematics}

In HEP the effect of systematic uncertainties is traditionally modeled using \textbf{template variations}, where the impact of each source is captured by a set of pre-defined shape variations (e.g.\ $\pm 1\sigma$ histograms) interpolated with a low-order polynomial in the nuisance parameters~\cite{CMS:2024combine,Cranmer:2012sba}. This procedure is robust and scales to hundreds of nuisance parameters in binned fits, but it operates on low-dimensional, binned summaries of the data and does not directly carry over to unbinned fits in high-dimensional feature spaces.

When the densities are modeled with generative models, the natural extension is to treat systematic uncertainties as \textbf{parametric deformations} of the learned probability densities. Implemented naively, however, this requires sampling the full $K$-dimensional space of nuisance configurations during training, whose cost grows exponentially with $K$. The \emph{Factorizable Normalizing Flow} (FNF) of Ref.~\cite{Valsecchi:2026fnf} was introduced precisely to solve this problem, and we adopt it here as the building block for the systematic-aware densities; the next subsection summarizes the construction and refers to that work for the full treatment.

Systematic uncertainties enter the fit in two distinct places, shown in Fig.~\ref{fig:concept_graph}. They first affect the \emph{input densities}, the reference model built from simulation: the shapes of $p_f(x\mid\nu)$ and $p_f(y\mid x,\nu)$ depend on the nuisances, a dependence that is \emph{known} and can be learned directly from simulated $\pm 1\sigma$ samples. They then affect the \emph{measurement} itself: as the nuisances vary, the best-fit Distribution of Interest must change to remain consistent with the data, a response that is \emph{not} available from simulation and is instead fixed by profiling the likelihood. We model both with the same Factorizable Normalizing Flow construction, through transformations that differ in the object they deform and in how their nuisance dependence is fixed: an \textbf{input systematic flow} $T_\chi$ acting on the reference densities (Section~\ref{sec:factorizable_flows}), and a \textbf{DoI systematic flow} $T_{\psi}$ acting on the measurement transformation (Section~\ref{sec:doi_systematics}).

\subsection{Factorizable Normalizing Flows for systematic uncertainties}
\label{sec:factorizable_flows}
To keep the paper self-contained we briefly recall the FNF construction, referring to Ref.~\cite{Valsecchi:2026fnf} for the full treatment and its validation. The idea is to separate the modeling of the nominal density from that of its systematic deformation. A fixed, high-fidelity normalizing flow $p_{\text{nom}}$ describes the data in the nominal configuration ($\nu = 0$), and a learnable, invertible transformation $T_\nu$ warps the feature space to absorb the effect of the nuisances (instantiated below as the input systematic flow $T_\chi$ and, in Section~\ref{sec:doi_systematics}, as the DoI systematic flow $T_\psi$). The systematic-aware density is the \emph{pullback} of the nominal one by $T_\nu$: an event is carried back to the nominal frame, scored under $p_{\text{nom}}$, and corrected by the Jacobian of the warp, which keeps the result a properly normalized density. The transformation is affine and autoregressive, acting on each feature as $y_{\mathrm{nom},j} = y_j\,e^{s_j} + t_j$, so that its Jacobian is triangular and cheap to evaluate. Its defining property is how the scale $s_j$ and shift $t_j$ depend on the nuisances: they are a low-order polynomial in $\nu$, a Taylor expansion around the nominal point, factorized over the individual nuisances,
\begin{equation}\label{eq:fnf_poly}
  s_j(y_{<j}, x, \nu) = \sum_{k=1}^{K}\Big(\nu_k\,\alpha_j^{k}(y_{<j}, x) + \nu_k^{2}\,\beta_j^{k}(y_{<j}, x)\Big) + \sum_{1 \le k < \ell \le K}\nu_k\,\nu_\ell\,\phi_j^{k\ell}(y_{<j}, x),
\end{equation}
and identically for the shift $t_j$ with its own coefficients. The per-nuisance coefficients $\alpha, \beta, \phi, \dots$ are produced by masked neural networks of the features, so the warp stays polynomial and smooth in $\nu$, as profiling requires, while remaining fully expressive in $(y, x)$. Because the dependence on $\nu$ is additive across the nuisances, each systematic can be learned in isolation from its own $\pm 1\sigma$ samples and the joint response of many nuisances is recovered at inference by summation, without ever sampling their combinatorially large joint space; the optional bilinear cross-terms $\nu_k\nu_\ell$ (coefficients $\phi$) restore the correlations between sources when these are not negligible (Appendix~\ref{sec:appendix_crossterms}). In this sense the FNF is the continuous, differentiable generalization of the $\pm 1\sigma$ template-variation recipe: the discrete up and down histograms become a smooth warp of the full density, and the per-bin polynomial interpolation becomes the polynomial dependence of that warp on $\nu$.

We first apply the FNF to the \emph{input} side of the fit: the systematic dependence of the reference densities, learned once from simulation. Fixed, high-fidelity nominal models $p_{\text{nom}, f}(x)$ and $p_{\text{nom}, f}(y \mid x)$, trained on high-statistics nominal simulation, are composed with learnable, invertible \emph{input systematic flows} $T_\chi^{x}$ and $T_\chi^{y}$ that pull each event back to the nominal reference. Since the systematics affect both the kinematic density $p_f(x\mid\nu)$ and the conditional density $p_f(y\mid x,\nu)$, each factor is the pullback of its nominal density,
\begin{equation}\label{eq:fnf_pullback}
\begin{aligned}
  p_f(x \mid \nu) &= p_{\text{nom}, f}\!\left(T_\chi^{x}(x \mid \nu)\right) \left|\det \nabla_{x} T_\chi^{x}(x \mid \nu)\right|, \\
  p_f(y \mid x, \nu) &= p_{\text{nom}, f}\!\left(T_\chi^{y}(y \mid x, \nu) \mid x\right) \left|\det \nabla_{y} T_\chi^{y}(y \mid x, \nu)\right|.
\end{aligned}
\end{equation}
Following the polynomial construction recalled above, both input flows $T_\chi^{x}$ and $T_\chi^{y}$ are affine and autoregressive, so the Jacobian is triangular and the likelihood remains tractable. Crucially, the coefficients of $T_\chi$ are calibrated on simulation, since the systematic effect on the inputs is known directly from the per-nuisance $\pm 1\sigma$ samples. In Section~\ref{sec:doi_systematics} we reuse the identical construction for the \emph{output} side, the Distribution of Interest, where the nuisance dependence is instead fixed by profiling the likelihood rather than read from simulation.

\subsection{Uncertainties on the Distribution of Interest}
\label{sec:doi_systematics}
The uncertainty on the inputs must be propagated to the final measurement. In a standard analysis this is the role of \emph{profiling}~\cite{Lista:2023stat,CMS:2024combine}: the likelihood $\mathcal{L}(\theta, \nu)$ is maximized jointly over the parameters of interest $\theta$ and the nuisances $\nu$, and the uncertainty on $\theta$ is read from the profile likelihood ratio $\lambda(\theta) = \mathcal{L}(\theta, \hat{\hat{\nu}}(\theta)) / \mathcal{L}(\hat{\theta}, \hat{\nu})$, with $\hat{\hat{\nu}}(\theta)$ the nuisances that maximize the likelihood at fixed $\theta$. The resulting interval on $\theta$ then includes the systematic variations allowed by the data and the nuisance constraints.

Two observations make the translation to our setting precise. First, profiling is fundamentally the partial maximization of the joint likelihood over the nuisances; scanning $\theta$ and recording the conditional optimum $\hat{\hat{\nu}}(\theta)$ is one organization of this operation, not part of its definition. Second, our parameter of interest is the transformation $T_{\phi}$, an entire distribution rather than a single number, which cannot be scanned point by point the way a scalar can. We therefore turn the procedure around and scan the nuisances instead: as $\nu$ varies over the region allowed by the data and its constraints, the best-fit transformation at each point, $\hat{T}_{\phi}(\nu)$, shifts accordingly, and the band it traces out is the systematic uncertainty on the measured distribution.

Concretely, profiling the Distribution of Interest amounts to optimizing the transformation and the nuisances together,
\begin{equation}\label{eq:profiling}
  \hat{T}_{\phi}, \hat{\nu} = \arg \max_{\phi, \nu} \mathcal{L}_{\text{ext}}(y, x \mid \nu, T_{\phi})
\end{equation}
where the factorizable structure of the input flow $T_\chi$ already propagates the systematic effects into the likelihood in a tractable way. Following the construction above, what profiling requires is this best-fit map $\hat{T}_{\phi}(\nu)$. Recovering it by re-optimizing the full transformation at each point of the nuisance space would be prohibitive; we instead make its nuisance dependence explicit.

We expand the transformation around its best-fit configuration $\hat{\nu}$: the nominal map $T_{\phi}^{\hat{\nu}}$, obtained at the best fit, is held fixed, while a second Factorizable Normalizing Flow, the \emph{DoI systematic flow} $T_{\psi}$, carries the entire variation with the nuisances. Mirroring the input side, $T_{\psi}$ acts on the measurement transformation rather than on the input densities, and is composed with the nominal DoI as
\begin{equation}\label{eq:Texp}
T_{\phi}(y | x, \Delta\nu) = T_{\phi}^{\hat{\nu}}(y | x) \circ T_{\psi}(y | x, \Delta \nu)
\end{equation}
Here $T_{\psi}$ uses the same factorizable construction as the input flow $T_\chi$~\cite{Valsecchi:2026fnf}, with one essential difference: its coefficients are not calibrated from simulation but are fixed by profiling the likelihood, so $T_{\psi}$ encodes how the measurement must deform to absorb each systematic variation. The joint optimization in Eq.~\ref{eq:profiling} then becomes:
\begin{equation}
  \hat{T}_{\phi}^{\hat{\nu}}, \hat{T}_{\psi}, \hat{\nu} = \arg \max_{\phi^{\hat{\nu}}, \psi, \nu} \mathcal{L}_{\text{ext}}(y, x | \nu, T_{\phi}^{\hat{\nu}} \circ T_{\psi})
\end{equation} 
Because $T_{\phi}(y | x, \nu)$ is built as a deviation from the nominal model, a small departure $\Delta\nu$ from the global minimum $\hat{\nu}$ keeps $T_{\psi}$ close to the identity; Section~\ref{sec:profiling-procedure} turns this into a tractable two-step training.

\subsection{Profiling systematics in unbinned likelihood fits}
\label{sec:profiling-procedure}

Optimizing the transformation and the nuisances together, as in Eq.~\ref{eq:profiling}, is difficult in practice: $T_{\phi}$ is high-dimensional, and reconstructing its dependence on the nuisances by re-optimizing it at every point of the nuisance space would be prohibitively expensive. We therefore split the profiling into two steps. In \textbf{Step 1}, a single global fit determines the best-fit nuisances $\hat{\nu}$ together with the nominal transformation $T_{\phi}^{\hat{\nu}}$, centering the measurement at the best-fit configuration; this is the first step of classical profiling. In \textbf{Step 2}, the nominal transformation is frozen and the residual systematic flow $T_{\psi}$ of Eq.~\ref{eq:Texp} is trained to capture how the Distribution of Interest must deform across the nuisance space, from which the systematic uncertainty on the measurement is obtained. The full procedure is given below and illustrated in Fig.~\ref{fig:amortized_profiling_detailed}; Step 2, the only computationally demanding part, is realized efficiently by an amortized training (Section~\ref{sec:amortized}).

\begin{figure}[!htbp]
    \centering
    \resizebox{\textwidth}{!}{%
\begin{tikzpicture}[
    node distance=1.5cm,
    font=\sffamily,
    layer/.style={rectangle, draw=black!80, fill=blue!10, very thick, minimum width=3cm, minimum height=1.5cm, rounded corners, align=center, drop shadow={opacity=0.15}},
    frozen/.style={rectangle, draw=black!50, fill=gray!10, very thick, dashed, minimum width=3cm, minimum height=1.5cm, rounded corners, align=center, drop shadow={opacity=0.15}},
    data/.style={rectangle, draw=black!80, fill=yellow!10, thick, minimum width=2cm, minimum height=1cm, align=center, drop shadow={opacity=0.15}},
    param/.style={circle, draw=black!80, fill=green!10, thick, minimum size=1.3cm, align=center, drop shadow={opacity=0.15}},
    loss/.style={rectangle, draw=black!80, fill=orange!20, thick, minimum width=2.5cm, minimum height=1.0cm, rounded corners, align=center, drop shadow={opacity=0.15}},
    connection/.style={->, >=stealth, thick, color=black!70},
    gradient/.style={->, >=stealth, very thick, color=red!80, dashed},
    transfer/.style={->, >=stealth, very thick, color=blue!60!black, dotted},
    step_label/.style={font=\bfseries\large, align=left, color=black!70}
]

\begin{scope}[yshift=7.5cm, local bounding box=step1]
    \node[step_label] at (-0.5, 3.5) {Step 1: Global Minimum\\ \small Joint Optimization};

    \node[param] (nu_param) at (0, 1.0) {$\nu$\\ \small params};
    \node[above=0.1cm of nu_param, left=0.1cm of nu_param, font=\scriptsize, color=red!80] {Learnable};

    \node[loss] (loss1) at (4.5, 1.0) {\textbf{Likelihood} $\mathcal{L}_{\text{ext}}(y, x | \nu)$\\ \small Maximize};

    \node[layer, drop shadow={opacity=0}] at (9.56, 1.36) {};
    \node[layer, drop shadow={opacity=0}] at (9.38, 1.18) {};
    \node[layer] (T_global) at (9.2, 1.0) {\textbf{Global Map} $T_{\phi}^{(b)}$ \\ \textit{Learnable} \\ {\scriptsize Poisson bootstrap, $b{=}1\dots K$}};

    \node[data, right=0.8cm of T_global] (data1) {\textbf{Batch Data}\\$(x, y)$};
    
    \draw[connection] (data1) -- (T_global);
    \draw[connection] (T_global) -- (loss1); 
    \draw[connection] (nu_param) -- (loss1); 

    \draw[gradient] (loss1.north) to[out=30, in=150] 
        node[pos=0.5, above, font=\scriptsize] {Update $\phi$} 
        (T_global.north);
        
    \draw[gradient] (loss1.north) to[out=150, in=30] 
        node[pos=0.5, above, font=\scriptsize] {Update $\nu$} 
        (nu_param.north);
\end{scope}

\begin{scope}[yshift=0cm, xshift=-1.9cm, local bounding box=step2]
    \node[step_label] at (1.5, 5.0) {Step 2: Systematic-Aware\\ \small Profile Training};

    \begin{scope}[local bounding box=nuisance_space]
        \node[font=\bfseries, align=center, scale=0.9] at (1.5, 4.2) {Nuisance Profile};
        \draw[->, thick] (0,0) -- (3.5,0) node[right] {$\nu_1$};
        \draw[->, thick] (0,0) -- (0,3.0) node[above] {$\nu_2$};
        
        \draw[color=gray!60, thick] (1.5,1.5) ellipse (1.5cm and 1.0cm);
        \draw[color=gray!80, thick] (1.5,1.5) ellipse (1.0cm and 0.6cm);

        \foreach \gx in {0.3,0.78,1.26,1.74,2.22,2.7}{
            \foreach \gy in {0.6,0.96,1.32,1.68,2.04,2.4}{
                \fill[red!55] (\gx,\gy) circle (1.4pt);
            }
        }
        \node[color=red!75, font=\scriptsize] at (1.5,2.95) {$\{\nu_q, w_q\}$};
        \coordinate (grid_out) at (2.7,2.2);

        \coordinate (best_fit_line) at (5, 5);
        \coordinate (best_fit) at (1.5, 1.5);
        \fill[green!60!black] (best_fit) circle (3pt);
        \node[font=\scriptsize, color=green!40!black, below right=-2pt and 2pt] at (best_fit) {$\hat{\nu}$};
    \end{scope}

    \node[data, below=1cm of nuisance_space] (data2) {\textbf{Batch Data}\\$(x, y)$};

    \node[layer, right=1.5cm of nuisance_space, yshift=-1.0cm] (T_sys) {\textbf{Systematic} $T_{\psi}$ \\ \textit{Learnable} \\ $y' = T_{\psi}(y | x, \nu_q)$};
    
    \node[frozen, drop shadow={opacity=0}] at ([shift={(2.86,0.36)}]T_sys.east) {};
    \node[frozen, drop shadow={opacity=0}] at ([shift={(2.68,0.18)}]T_sys.east) {};
    \node[frozen, right=1.0cm of T_sys] (T_nom) {\textbf{Nominal} $T_{\phi}^{(b)}$ \\ \textit{Frozen member} \\ $z = T_{\phi}^{(b)}(y' | x)$};

    \node[loss, below=1.0cm of T_nom] (loss2) {\textbf{Likelihood} $\sum_q w_q\,\mathcal{L}_{\text{ext}}(y, x| \nu_q)$};

    \draw[connection, color=red!80, very thick] (grid_out) -- ++(3,0) node[midway, above, font=\scriptsize, color=red] {quadrature grid} -| (T_sys.north);
    \draw[connection] (data2) -- ++(3.5,0) |- (T_sys.west);
    \draw[connection] (T_sys) -- node[midway, above, font=\scriptsize] {$y'$} (T_nom);
    \draw[connection] (T_nom) -- (loss2.north);
    \draw[gradient] (loss2.west) -| node[pos=0.75, left=0.1cm, align=right, font=\scriptsize] {\textbf{Update}\\$\psi$ only} (T_sys.south);
\end{scope}

\draw[transfer] (nu_param.south) to[out=-90, in=90] (best_fit_line) to[out=-90, in=90] (best_fit);
\node[right, align=right, font=\small, text width=2.5cm, fill=none, inner sep=5pt, rounded corners, color=blue!60!black] at (best_fit_line) {Fix Best-Fit $\hat{\nu}$\\as Profile Center};

\draw[transfer] (T_global.south) -- (T_nom.north)
    node[midway, right, align=left, font=\small, text width=3.0cm, fill=none, inner sep=4pt, rounded corners, xshift=0.2cm] {Freeze ensemble;\\draw member $T_{\phi}^{(b)}$\\each step ($T_{\psi}$ shared)};

\begin{pgfonlayer}{background}
    \node[draw=gray!30, fill=white, very thick, rounded corners, fit=(step1), inner sep=0.5cm, drop shadow={opacity=0.1}] {};
    
    \node[draw=gray!30, fill=white, very thick, rounded corners, fit=(step2) (data1.east |- step2.east), inner sep=0.5cm, drop shadow={opacity=0.1}] {};
\end{pgfonlayer}

\end{tikzpicture}
    }%
    \caption{Schematic of the proposed training procedure. First a global minimum is found by jointly optimizing the nominal transformation $T_{\phi}$ and nuisance parameters $\nu$; this Step 1 fit is repeated as a Poisson bootstrap, producing the ensemble $\{T_{\phi}^{(b)}\}$ of nominal maps (stacked boxes) that carries the statistical uncertainty of the measurement. Then, the amortized training for systematics profiling starts. Rather than at a single point, the residual is optimized on a fixed Gauss-Legendre quadrature grid $\{\nu_q, w_q\}$ covering the data-constrained nuisance region delimited by the confidence contours of the global fit (top left). The "Systematic Map" network $T_{\psi}$ is evaluated at every grid node $\nu_q$ (conditioning) along with the data, and is shared across the ensemble, a random member $T_{\phi}^{(b)}$ being drawn and frozen at each step. Each gradient descent step updates the network parameters to maximize the weighted sum of the per-node log-likelihoods, $\sum_q w_q \ln \mathcal{L}_{\text{ext}}(\nu_q)$, so the network learns the conditionally optimal deformation at every node at once and, through its low-order polynomial dependence on $\nu$, interpolates between them, approximating the profiled map across the entire nuisance space in a single training.}
    \label{fig:amortized_profiling_detailed}
\end{figure}

\subsubsection{The two-step procedure}

The procedure consists of three steps:
\begin{enumerate}
  \item \textbf{Global fit (Step 1).} Jointly optimize the transformation and the nuisances on the full dataset,
  \begin{equation}
    \hat{T}_{\phi}^{\hat{\nu}}, \hat{\nu} = \arg\max_{\phi, \nu} \mathcal{L}_{\text{ext}}(y, x \mid \nu, T_{\phi}),
  \end{equation}
  which fixes the best-fit nuisances $\hat{\nu}$ and the corresponding best-fit transformation $T_{\phi}^{\hat{\nu}}$. We stress that $T_{\phi}^{\hat{\nu}}$ is a single, fixed map with \emph{no} explicit dependence on $\nu$: the superscript $\hat{\nu}$ only labels the anchor at which the map was obtained, not a functional argument, and it enters the composition of Eq.~\ref{eq:Texp} as $T_{\phi}^{\hat{\nu}}(y\mid x)$ alone. For clarity of exposition, we describe the procedure here for a single measurement map. In practice, to capture the statistical uncertainty of the fit, this step is repeated over a Poisson-bootstrap to form an ensemble of nominal maps, as detailed in Section~\ref{sec:stat_uncertainty}. 
  \item \textbf{Decomposition.} Freeze the nominal transformation and write the full Distribution of Interest as the composition $T_{\phi}^{\hat{\nu}} \circ T_{\psi}$ of Eq.~\ref{eq:Texp}, leaving only the parameters of the residual flow $T_{\psi}(\nu)$~\cite{Valsecchi:2026fnf} free.
  \item \textbf{Systematic-aware training (Step 2).} With the nominal transformation frozen, train the residual flow to give, at each nuisance configuration, the deformation that maximizes the likelihood,
  \begin{equation}
    \hat{T}_{\psi}(\nu) = \arg\max_{\psi} \mathcal{L}_{\text{ext}}(y, x \mid \nu, T_{\phi}^{\hat{\nu}} \circ T_{\psi}),
  \end{equation}
  so that the full transformation becomes a function of the nuisances, $\hat{T}_{\phi}(\nu) = T_{\phi}^{\hat{\nu}} \circ \hat{T}_{\psi}(\nu)$. When deploying the bootstrap ensemble (Section~\ref{sec:stat_uncertainty}), this single residual flow $T_{\psi}$ is trained once and shared across all members of the ensemble.
\end{enumerate}

The systematic uncertainty on the Distribution of Interest is then read off by evaluating $\hat{T}_{\phi}(\nu)$ along the likelihood contour determined by the global fit, with frequentist coverage (see \ref{sec:results:coverage}) and the contribution of each systematic kept interpretable. Step 2 is the only demanding part, since it requires the residual flow across the whole nuisance space rather than at a single point; we realize it efficiently with an amortized training, described next.

\subsubsection{Amortized profiling}
\label{sec:amortized}

Obtaining $\hat{T}_{\psi}(\nu)$ by a separate fit at each point of the nuisance space would be prohibitively expensive. We instead train $T_{\psi}$ in an \textbf{amortized} fashion: because it is conditioned on $\nu$, a single network represents the residual for every nuisance configuration. During training we draw $\nu$ uniformly over an interval wide enough to cover the range scanned at inference, so that the residual is determined everywhere it will be evaluated, and maximize the resulting average of the extended log-likelihood,
\begin{equation}\label{eq:amortized_objective}
  \hat{T}_{\psi} = \arg\max_{\psi}\ \mathbb{E}_{\nu \sim \mathcal{U}}\!\left[\ln \mathcal{L}_{\text{ext}}(y, x \mid \nu, T_{\phi}^{\hat{\nu}} \circ T_{\psi})\right],
\end{equation}
with $\mathcal{U}$ the uniform distribution over that interval. This interval is a modelling choice. Because the residual is a local, low-order expansion of the best-fit transformation around the anchor $\hat{\nu}$ (Section~\ref{sec:doi_systematics}), the range must be wide enough to contain the scanned region yet narrow enough that this expansion stays accurate; in practice we match it to the data-constrained region, the confidence region of the global fit, and a substantially wider range would require a higher polynomial degree.

Estimating this average with a single $\nu$ drawn per event is unbiased but high in variance, and the noise falls precisely on the higher-order (quadratic and cross) terms of the residual that carry the correlations between nuisances. We instead evaluate the average on a \textbf{Gauss-Legendre quadrature grid} $\{\nu_q, w_q\}$~\cite{Golub:1969quad,Press:2007nr},
\begin{equation}\label{eq:amortized_quadrature}
  \hat{T}_{\psi} = \arg\max_{\psi}\ \sum_{q} w_q\, \ln \mathcal{L}_{\text{ext}}(y, x \mid \nu_q, T_{\phi}^{\hat{\nu}} \circ T_{\psi}),
\end{equation}
which, with $Q$ nodes per nuisance, integrates polynomials up to degree $2Q-1$ exactly and converges far faster than random sampling for smooth integrands.

A fixed grid is appropriate here, rather than merely convenient, because the nuisance dependence of the residual is a \emph{fixed low-order polynomial}: as in the input flow (Section~\ref{sec:factorizable_flows}), the network produces only the $x$-dependent coefficients while $\nu$ enters the scale and shift analytically. Crucially, while the integrand $\ln \mathcal{L}_{\text{ext}}$ is highly non-linear and not a polynomial in $\nu$, this does not break the exactness of the grid. By choosing $Q \geq p+1$ nodes per nuisance, the mapping between the $p+1$ predicted network coefficients and the transformation values at the $Q$ nodes is an exact bijection. Consequently, maximizing the sum over the grid is mathematically equivalent to independently maximizing the likelihood at each of the $Q$ nodes. The network perfectly identifies the true, conditionally optimal transformation at each node regardless of the shape of the likelihood function, and the polynomial ansatz simply interpolates between these optimal maps. The $d$-dimensional grid is the tensor product of the one-dimensional rules at a cost of $Q^{d}$ evaluations per event. A single training run thus learns the systematic response, with its nuisance correlations, over the entire nuisance space, turning the profiling scan into a one-time training cost. An example of the response that $T_{\psi}$ learns is shown in Appendix~\ref{sec:appendix_residual}.

The tensor-product cost $Q^{d}$ makes the quadrature most effective at \emph{low dimensionality}. In the two-nuisance validation of Section~\ref{sec:experiments} a handful of nodes per axis integrate the low-order residual exactly at negligible cost, and for the few-nuisance analyses typical of a focused measurement the grid remains the natural choice: it is deterministic, noise-free, and exact for the polynomial residual. Its cost, however, grows exponentially with the number of nuisances, the usual curse of dimensionality of product rules, so beyond a handful of parameters a dense grid quickly becomes impractical. In that regime the deterministic grid is better replaced by a \emph{stochastic} estimator of the same expectation in Eq.~\ref{eq:amortized_objective}: Monte-Carlo sampling of $\nu$, whose error decreases as $1/\sqrt{N}$ independently of $d$. The single-$\nu$-per-event estimator discussed above is the simplest such scheme; coupled with variance-reduction techniques to temper the high variance noted there, it would retain this dimension-independent scaling. The choice between quadrature and sampling is therefore governed by the nuisance count: a fixed grid for the few-nuisance setting demonstrated here, and Monte-Carlo integration for the many-nuisance regime of a realistic analysis.

\subsubsection{Orthogonal decomposition of the uncertainty space}
\label{sec:orthogonal_decomposition}
The systematic band obtained above captures the full effect of the nuisances on the Distribution of Interest, but with many correlated parameters it is hard to see which directions actually drive it. We therefore decompose the nuisance confidence region, the local description of which is the Hessian of the likelihood at the best fit, into orthogonal modes, and read off the effect of each mode on the measurement by evaluating $T_{\psi}$ along it.

Concretely, the decomposition proceeds in three steps:
\begin{itemize}
    \item \textbf{Hessian Analysis}: We compute the Hessian matrix of the negative log-likelihood with respect to $\nu$ at the global minimum $\hat{\nu}$ using automatic differentiation. This matrix captures the local curvature and correlations between nuisance parameters. See Appendix~\ref{sec:appendix_bma_hessian} for details.
    \item \textbf{Eigen-decomposition}: The eigendecomposition of the Hessian yields orthogonal eigenvectors representing independent modes of uncertainty, with eigenvalues quantifying their constraining power.
    \item \textbf{Principal Mode Visualization}: We visualize the ``principal systematic variations'' by evaluating the trained map $T_{\psi}$ along these eigenvector directions. This reveals the dominant shapes of uncertainty driving the analysis, decoupled from the original (potentially correlated) physics parametrization.
\end{itemize}
  
This procedure is showcased in Fig.~\ref{fig:orthogonal_decomposition}, where we visualize the principal systematic variations obtained from the orthogonal decomposition of the uncertainty space on the synthetic dataset. The plot displays the effect of the leading eigenvector on the DoI transformation, providing insights into the dominant mode of systematic variation and its impact on the measurement.


\subsection{Statistical uncertainty of the Distribution of Interest}
\label{sec:stat_uncertainty}

The profiling of Section~\ref{sec:profiling-procedure} propagates the \emph{systematic} uncertainty, the effect of the nuisances, but the measurement carries a second, statistical component. The nominal transformation $T_{\phi}$ is the result of a maximum-likelihood fit to a finite dataset, and like any estimator it would change if the measurement were repeated on statistically independent data. Since the transformation is itself the measurement, this finite-sample variability is a genuine statistical uncertainty on the Distribution of Interest, and it must be propagated into the inference alongside the systematic one.

We estimate it with a \textbf{Poisson bootstrap} of the Step 1 fit. We repeat the fit $K$ times on the same dataset, each time weighting every event by an independent $w_i \sim \mathrm{Poisson}(1)$, which reproduces the multinomial bootstrap in expectation, and share the initialization and data ordering so that the only difference between members is the set of weights. The resulting ensemble $\{T_{\phi}^{(b)}\}_{b=1}^{K}$ samples the transformations compatible with the data, and its spread is the data-statistical uncertainty on the measurement.

The systematic residual $T_{\psi}$ is trained once and shared across the ensemble: at each step a random member $T_{\phi}^{(b)}$ is composed with it, so $T_{\psi}$ learns the ensemble-averaged nuisance response around a single common anchor rather than being refit for each member. The statistical uncertainty therefore resides entirely in the ensemble, while the systematic response is described by a single residual.

At inference the two contributions are combined by evaluating the profiled likelihood $L_b(\nu)$ of every member and averaging across the ensemble. The members are not competing hypotheses or alternative fit models, as in a discrete-template or classical envelope construction where one profiles over the alternatives and keeps the best at each point; they are bootstrap replicas of a single measurement, sampling the statistical fluctuation of $T_{\phi}$. Propagating that fluctuation means marginalizing the likelihood over it, that is averaging the per-member likelihoods rather than taking their envelope,
\begin{equation}\label{eq:bma}
  -2\ln L_{\mathrm{ens}}(\nu) = -2\ln\!\left[\frac{1}{K}\sum_{b=1}^{K} L_b(\nu)\right].
\end{equation}
We refer to this equal-weight combination as the \textbf{ensemble-averaged}, or \textbf{bagged}, likelihood. The members enter with uniform weight $1/K$, with no model prior and no evidence weighting, so the sum is a Monte-Carlo estimate of the likelihood marginalized over the bootstrap sampling distribution of $\hat{T}_{\phi}$, $\frac{1}{K}\sum_b L_b(\nu) \to \mathbb{E}_{T_{\phi}}\!\left[L(\nu \mid T_{\phi})\right]$. It is the bootstrap aggregation (bagging) of the likelihood~\cite{breiman1996bagging}: each replica is a fit to a Poisson-resampled copy of the data, and averaging their likelihoods marginalizes the finite-sample variability of the fitted transformation into the inference~\cite{lakshminarayanan2017simple,newton1994approximate}, in the same spirit as the hybrid marginalization of nuisance parameters familiar from HEP. The content is therefore frequentist; the equal-weight average would coincide with a Bayesian model average only in the degenerate case of equal model evidence, which is not the mechanism at work here. Averaging the likelihoods broadens the combined curve by the spread of the per-member fits, so the inferred nuisance interval absorbs the statistical uncertainty of $T_{\phi}$ on top of the systematic profiling.

A subtlety is specific to unbinned fits: the absolute log-likelihood at the optimum varies substantially from member to member, a per-event difference of the order of the training noise becoming a large, $\nu$-independent offset in $-2\ln L$ once summed over the full event sample. Such an offset is pure normalization, which frequentist inference reads through the likelihood ratio and never through the absolute likelihood; left in, it dominates the exponential in Eq.~\ref{eq:bma} and collapses the average onto whichever replica sits lowest, discarding the ensemble spread. We therefore average the per-member likelihood ratios, rebasing each member to its own optimum, $-2\Delta\ln L_b(\nu) = -2\ln L_b(\nu) - \min_{\nu}\!\left[-2\ln L_b(\nu)\right]$. Subtracting this single scalar cannot alter the $\nu$-dependence of any member, only the uninformative weight between them, which the bootstrap fixes to be equal; the effective number of contributing members confirms that the average does not collapse onto one. The combined profile is then itself referred to its own minimum,
\begin{equation}\label{eq:bma_rebased}
  -2\Delta\ln L_{\mathrm{ens}}(\nu) = -2\ln L_{\mathrm{ens}}(\nu) - \min_{\nu}\!\left[-2\ln L_{\mathrm{ens}}(\nu)\right],
\end{equation}
so that its best fit sits at zero and the confidence contours follow from the usual $\Delta\chi^2$ thresholds. This second rebasing is needed because the per-member optima lie at different points in the nuisance plane, so the average of the rebased members does not in general reach zero at any single point. Both rebasings, of the individual members and of their average, are applied to the scans shown in Figures~\ref{fig:likelihood_scans} and~\ref{fig:likelihood_scans_step2}.

\subsection{Training dynamics}
\label{sec:training_dynamics}
The Step 1 fit optimizes the high-dimensional transformation $\phi$ and the few nuisances $\nu$ jointly, two sets of parameters that live at very different scales. As in any such joint optimization, balancing their updates calls for some care to ensure smooth and reproducible convergence, which a few standard techniques readily provide:
\begin{itemize}
  \item \textbf{Alternating optimization}: fix one set of parameters while updating the other, alternating between $\phi$ and $\nu$, which damps oscillations between them.
  \item \textbf{Learning-rate scheduling}: use separate learning rates for $\phi$ and $\nu$, for instance a larger rate for the few nuisances and a smaller one for the high-dimensional transformation.
  \item \textbf{Gradient accumulation on the nuisances}: accumulate and average the nuisance gradient over several batches before each update of $\nu$, which reduces its stochastic noise and allows a larger learning rate for $\nu$ to be used without destabilizing the optimization.
  \item \textbf{Regularization}: penalize large updates of $\phi$ or $\nu$ to encourage smoother, more stable steps.
\end{itemize}

A separate difficulty arises if the objective has a degenerate direction in the nuisances: as in classical profiling, a nuisance that the data do not constrain leaves the fit ill-posed. Such degeneracies are broken by the constraint terms already present in the likelihood, or by further information such as control regions or auxiliary measurements that constrain the nuisances independently.

\subsection{Comparison with related work}
\label{sec:related_work}
Several directions of research are exploring SBI methods for unbinned likelihood fits in HEP, with a particular focus on building multidimensional likelihood ratios using classifiers trained on simulated data \cite{Cranmer:2020wdu,Butter:2025dm,ATLAS:2025nsbi,Schofbeck:2025smeft,Barrue:2026qul,Szewc:2025hisigma,Brehmer:2018eca,T2K:2025omnifold,Andreassen_2020,GomezAmbrosio:2022step,Amram:2025hisigma}. These methods typically involve training a classifier to distinguish between different hypotheses (e.g., signal vs background) and then using the output of the classifier to construct a likelihood ratio that can be used for inference. A recent milestone in this direction is the ATLAS measurement of off-shell Higgs boson production in the $H^* \rightarrow Z Z \rightarrow 4\ell$ decay channel~\cite{ATLAS:2025higgs,ATLAS:2025nsbi}, which pioneers the application of neural simulation-based inference in a full-scale LHC analysis to perform an unbinned likelihood fit directly in a high-dimensional feature space. 
This analysis represents a significant milestone in the application of SBI methods for unbinned likelihood fits in HEP, demonstrating the potential of these techniques to improve the precision and robustness of measurements in complex analyses. However, the handling of the systematic uncertainties in this analysis required a significant computational effort, as the effect of each nuisance parameter was modeled as a vertical interpolation of the likelihood using additional classifiers trained on the $\pm 1 \sigma$ variations of each nuisance parameter. This approach, while effective, is computationally intensive and does not scale well with the number of nuisance parameters, as it requires training a separate classifier for each variation. Finally, the Hessian matrix of the likelihood at the best-fit point was used only to derive the pulls and impacts of the nuisance parameters, exploiting the automatic differentiation capabilities of the framework, while the uncertainty on the parameter of interest (POI) was obtained through a Neyman construction. The training of these classifiers on finite simulation is itself a source of statistical uncertainty. While this is addressed through an ensemble of bootstrapped networks, the resulting variance on the parameter estimate is evaluated separately on Asimov datasets and then injected back into the fit as an additional nuisance parameter with a Gaussian constraint, rather than being marginalized directly within the primary likelihood contour used for the Neyman construction.

Another similar direction of research is the parametric modelling of systematic uncertainties, as proposed in \cite{Schofbeck:2025smeft,zwzt-1rrw} and recently applied to the extraction of proton structure at the LHC~\cite{Barrue:2026qul}. This approach assumes that the effect of a systematic variation on the data can be factorized from the physics parameter of interest $\theta$, such that:
\begin{equation}
p(x \mid \theta, \nu) =  p(x \mid \theta, \nu_0) \cdot \frac{p(x \mid \nu)}{ p(x \mid \nu_0)}.
\end{equation}
To avoid retraining models for every variation, the systematic likelihood ratio $p(x \mid \nu) / p(x \mid \nu_0)$ is estimated via the ``likelihood ratio trick.'' A classifier $\hat{f}(x)$ is trained to separate nominal from varied datasets, parameterized as a logistic sigmoid:
\begin{equation}
\hat{f}(x) = \frac{1}{1 + \exp(-\hat{g}(x))}
\end{equation}
where the log-odds function $\hat{g}(x)$ is explicitly expanded as a polynomial in the nuisance parameters around the nominal configuration $\nu_0$:
\begin{equation}
\hat{g}(x; \nu) = \nu^T \hat{\Delta}_1(x) + \nu^T \hat{\Delta}_2(x) \nu + \dots
\end{equation}
The coefficient functions $\hat{\Delta}_k(x)$ are implemented as neural networks or boosted-decision trees that capture the kinematic dependence of the systematic effects. 
While this refinable ansatz provides a highly efficient way to evaluate systematic variations continuously, its fundamental limitation lies in the factorized assumption: it rigidly assumes that the effect of the systematic uncertainties is entirely independent of the parameter of interest. It therefore cannot handle conditional densities where the physics parameters and the detector systematics are deeply coupled, a limitation our framework naturally overcomes by profiling the full non-factorizable transformation.

Our approach expands the current landscape of SBI methods in three directions: first, we expand the target of the measurement to a multidimensional Distribution of Interest, represented by an invertible transformation, instead of a single POI $\theta$. This allows us to capture more complex and high-dimensional measurements, which are becoming increasingly common in HEP analyses. Second, we introduce a novel method for modeling systematic uncertainties that allows us to learn the effect of systematic variations on the likelihood across the entire nuisance parameter space in a single training run, leveraging the factorizable structure of the transformations. This amortized training procedure significantly reduces the computational cost of profiling systematic uncertainties, making it feasible to handle a large number of nuisance parameters in unbinned likelihood fits. Third, we propagate the finite-sample statistical uncertainty of the learned model as a first-class component of the measurement. The neural network that defines the Distribution of Interest is fitted to a finite dataset and therefore carries a statistical error of its own; we sample this variability with a Poisson-bootstrap ensemble of the fit and fold it into the inference through an ensemble-averaged (bagged) likelihood, so that the statistical and systematic uncertainties are combined within a single unbinned likelihood. This contrasts with existing classifier-based approaches. While recent works like the ATLAS analysis discussed above do utilize ensembles of networks to evaluate the training statistical uncertainty, they do so by performing separate fits across the ensemble to extract a variance on the fitted parameters, which is then added back to the main likelihood as an effective Gaussian nuisance parameter. In our approach, we fold this finite-sample variability directly into the inference through an ensemble-averaged (bagged) likelihood. As a result, the training statistical uncertainty is naturally marginalized and combined with the systematic profiling within a single, unified unbinned likelihood contour, without requiring an external variance calculation or ad-hoc auxiliary constraints. Taken together, these three directions deliver a complete statistical-plus-systematic uncertainty budget for a functional, neural-network measurement.

Targeting a Distribution of Interest instead of a single POI is also a direction explored in the literature, with methods such as morphing-based approaches \cite{Algren:2025gap,ATLAS:2025calib} that aim to learn optimal mapping between data and simulation using Optimal Transport techniques~\cite{amos2017inputconvexneuralnetworks,Peyre:2019cot}. Although these methods provide a powerful way to capture complex distributions with mathematical guarantees on the optimality of the mapping, they are not designed to handle systematic uncertainties in a way that allows for profiling, as they typically focus on learning a single mapping between data and simulation without explicitly modeling the dependence on nuisance parameters, and, being point estimates of that mapping, they do not quantify the finite-sample statistical uncertainty of the learned transformation either. Hence, including systematic variations in these approaches would require training separate morphing functions for each variation, which can be computationally expensive and may not capture the full complexity of the likelihood landscape, especially in high-dimensional feature spaces~\cite{ATLAS:2025calib}. Using Normalizing Flows to learn the morphing~\cite{Daumann:2024flow} permits instead to compute the likelihood efficiently and implement the profiling procedure in a tractable way. Because they are learned inside a likelihood, these transformations can themselves be interpreted as approximate Optimal Transport maps between the reference and the data conditioned on the nuisances, and an explicit transport map can be recovered as a post-processing step, for instance with Conditional Flow-Matching~\cite{Tong:2024otcfm} or Input Convex Neural Networks~\cite{huang2021convexpotentialflowsuniversal}; we leave the exploration of this aspect to future work.

\section{Experiments}
\label{sec:experiments}

To validate the proposed framework and its associated training procedures, we conduct a series of experiments on a synthetic dataset designed to mimic common scenarios in HEP analyses. These experiments aim to demonstrate the effectiveness of FNF in accurately modeling systematic uncertainties and improving the robustness of unbinned likelihood fits.

\subsection{Synthetic dataset}
\label{sec:dataset}

\begin{figure}[htb]
  \centering
  \includegraphics[width=\textwidth]{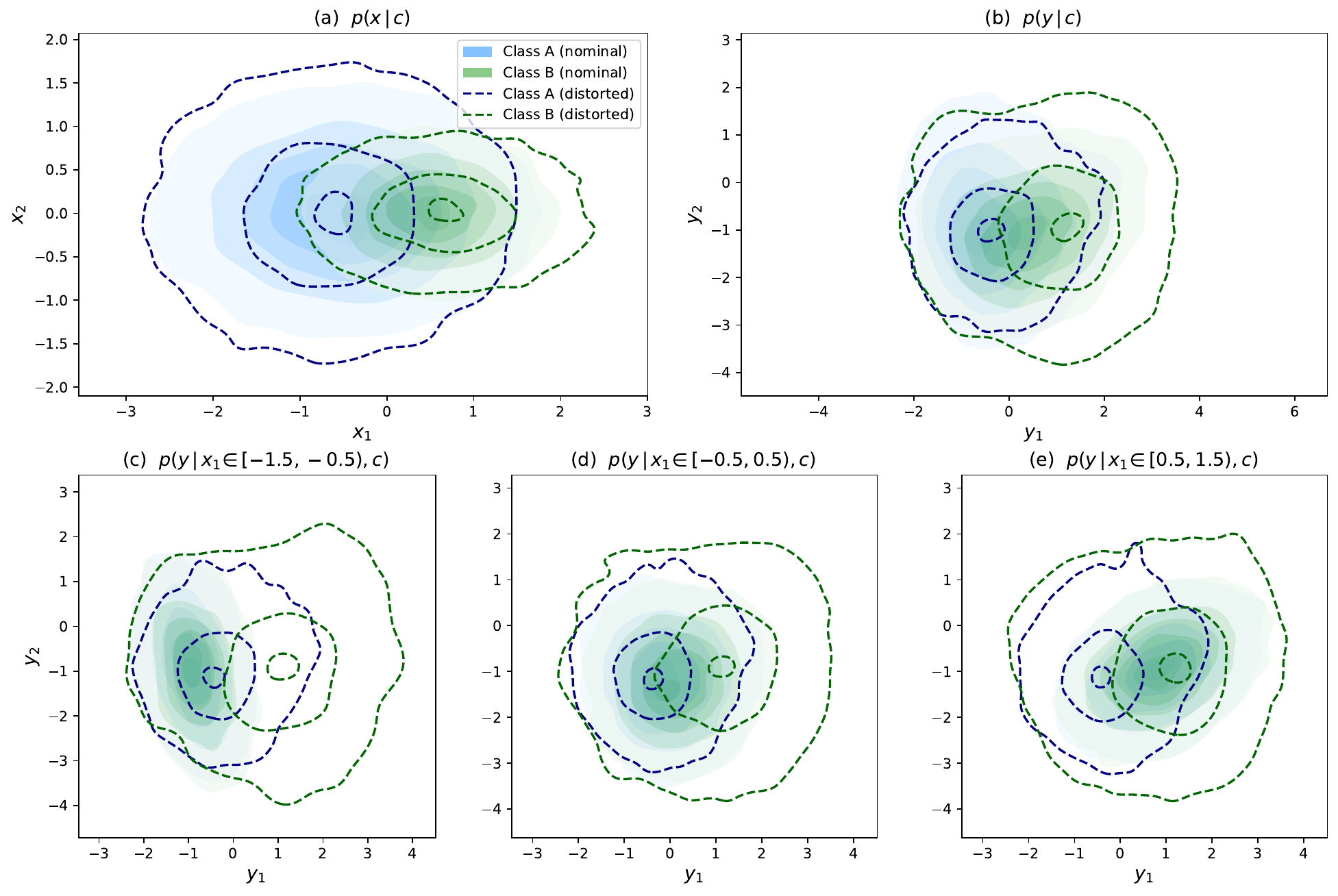}
  \caption{Structure of the synthetic dataset, comparing the nominal model (filled contours) with the distorted pseudo-data (dashed contours) for class A (blue) and class B (green). \emph{(a)} kinematic density $p_f(x)$; \emph{(b)} marginal score density $p_f(y)$; \emph{(c, d, e)} conditional score density $p_f(y \mid x)$ in three slices of $x_1$, illustrating its non-trivial dependence on the conditioning kinematics. The pseudo-data carry the distortion on the score together with the nuisances at their injected values, so they depart from the nominal model both in the kinematic and in the score distributions.}
  \label{fig:dataset}
\end{figure}

We validate the method on the controlled synthetic dataset introduced in Ref.~\cite{Valsecchi:2026fnf} to study the Factorizable Normalizing Flows, here extended with an explicit data-simulation distortion that serves as the measurement target. In a low-dimensional, fully controlled setting it reproduces the essential ingredients of a systematic-aware unbinned fit: a non-trivial nominal density, two interpretable systematic deformations, and a data-simulation mismodelling that the Distribution of Interest must recover.

Each event carries a binary class label $f \in \{A, B\}$, drawn with equal probability, a two-dimensional kinematic vector $x = (x_1, x_2)$, and a two-dimensional feature vector $y = (y_1, y_2)$, the ``score''. The pair $(f, x)$ provides the conditioning information, while $y$ is the observable whose density we model. The kinematics follow two class-dependent Gaussian clusters,
\begin{equation}
x \mid f \sim \mathcal{N}\!\big(\mu_f,\ \mathrm{diag}(\sigma_f^2)\big), \qquad
\begin{aligned}
\mu_A &= (-0.5,\, 0), & \sigma_A &= (0.9,\, 0.6),\\
\mu_B &= (+0.5,\, 0), & \sigma_B &= (0.6,\, 0.4),
\end{aligned}
\end{equation}
while the nominal feature density $p_f(y \mid x)$ is a bivariate Gaussian whose mean, per-axis spread, and correlation depend non-trivially on the kinematics (Appendix~\ref{sec:appendix_dataset}). This nominal configuration, with the nuisances at zero and no distortion applied, defines the reference simulation on which the input density flows are trained. Figure~\ref{fig:dataset} shows the kinematic and score densities for the two classes, together with the effect of the distortion introduced below.

\paragraph{Data-simulation distortion.}
The pseudo-data are drawn from the same generator and then passed through a fixed \textbf{distortion map} $F(y \mid x, f)$ acting on the score, which stands in for the data-simulation mismodelling that a real analysis must correct. The map composes a class-dependent rotation of the score, whose magnitude decays with the kinematic radius $|x|$ and is antisymmetric in $x_1$, with a small class-dependent shear that mixes the two score components, the two classes being deformed with opposite sign. The distortion is genuinely $x$-dependent and cannot be reproduced by any configuration of the nuisances, so it is captured only by the Distribution of Interest $T_{\phi}$: recovering $F$ from the data is the measurement. Its explicit form is given in Appendix~\ref{sec:appendix_dataset}.

\paragraph{Systematic variations.}

\begin{figure}[hbtp]
    \centering
    \includegraphics[width=\textwidth]{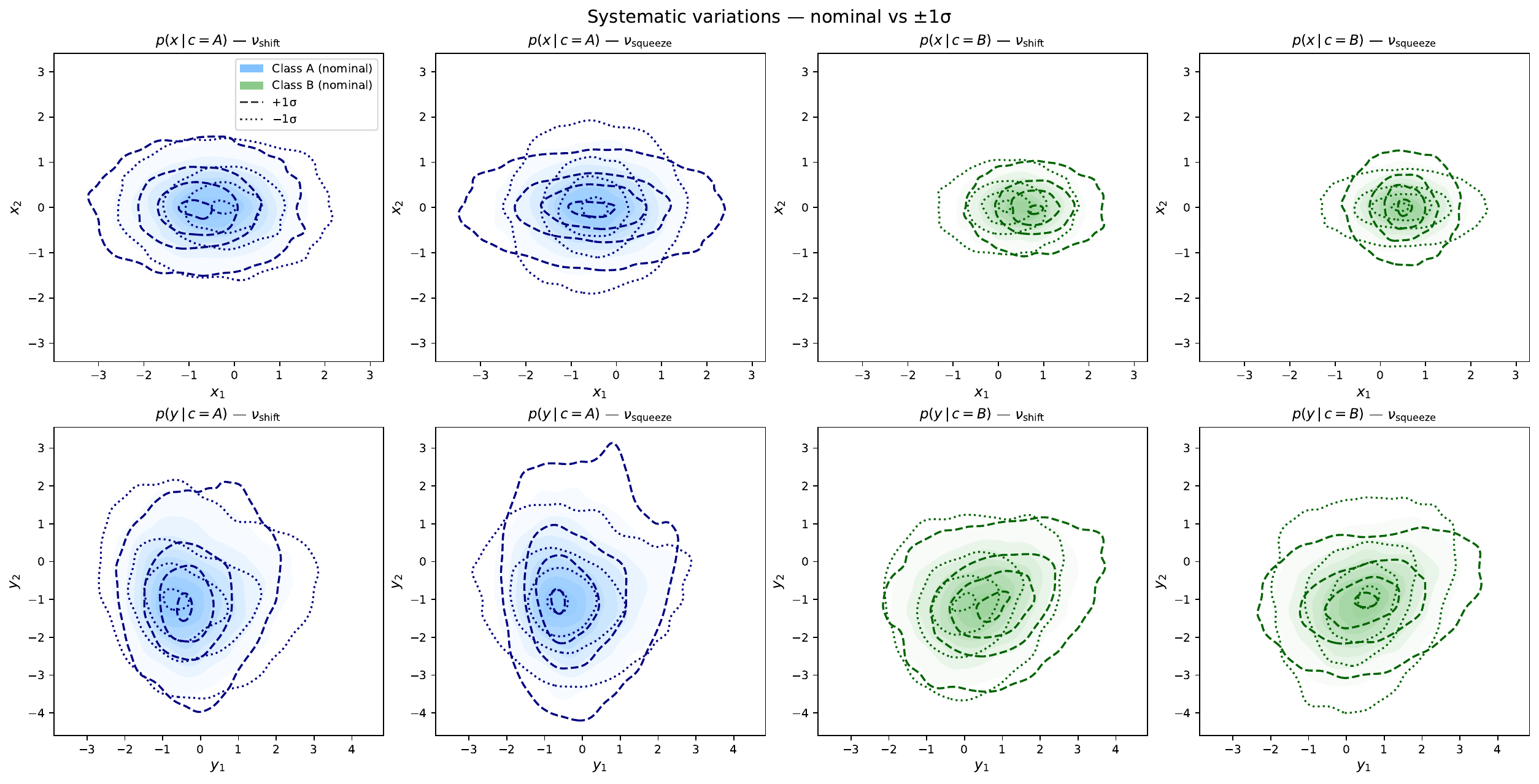}
    \caption{Effect of the two nuisance parameters on the input densities, comparing the nominal model ($\nu = 0$, filled contours) with the $+1\sigma$ (dashed) and $-1\sigma$ (dotted) variations. \emph{Top:} kinematic density $p_f(x)$; \emph{bottom:} marginal score density $p_f(y)$. The four columns give, for class A (blue) and class B (green), the response to $\nu_{\text{shift}}$ and to $\nu_{\text{squeeze}}$ in turn. Each nuisance acts with opposite sign on the two classes and deforms both the kinematics and the score; these are the variations the input flows learn from the per-nuisance $\pm 1\sigma$ simulation.}
  \label{fig:nuisance_xy}
\end{figure}

Two nuisance parameters, $\nu = (\nu_{\text{shift}}, \nu_{\text{squeeze}})$, drive the systematic uncertainties. Each acts simultaneously on the kinematics and on the conditional score, with opposite sign on the two classes:
\begin{itemize}
  \item $\nu_{\text{shift}}$ translates the cluster centroids along $x_1$, $x \to x \mp \nu_{\text{shift}}\, s_{\text{shift}}\, \hat{d}$ (with $\hat{d} = (1, 0)$; $-$ for class $A$ and $+$ for class $B$), and adds a linear, $x$-dependent shift to the score mean;
  \item $\nu_{\text{squeeze}}$ applies a volume-preserving ($\det = 1$), axis-anti-correlated squeeze of the kinematics about each centroid, $x \to \mu_f + \mathrm{diag}(e^{+\alpha}, e^{-\alpha})\,(x - \mu_f)$ with $\alpha = \nu_{\text{squeeze}}\, s_{\text{squeeze}}$ (the sign mirrored between classes), together with an exponential rescaling of the score spread.
\end{itemize}
Because both nuisances perturb the conditioning kinematics \emph{and} the score, they induce a genuine, interpretable dependence of $p_f(x \mid \nu)$ and $p_f(y \mid x, \nu)$ on the nuisances, which the input flows learn from the per-nuisance $\pm 1\sigma$ simulation. The kinematic deformation scales are $s_{\text{shift}} = 0.3$ and $s_{\text{squeeze}} = 0.2$, with feature-space response scales of $0.3$ and $0.2$ respectively. Figure~\ref{fig:nuisance_xy} shows their effect, at the $\pm 1\sigma$ variations, on the kinematic density $p_f(x)$ and on the score density $p_f(y)$.

\paragraph{What the fit measures.}
The two effects are distinct by construction, which is what makes the dataset a meaningful test of profiling. The distortion $F$ is common to every nuisance configuration and is therefore owned by the measurement, the DoI $T_{\phi}$; the nuisances instead reshape the reference densities, and the way their variation displaces the best-fit transformation is exactly what the profiling of Section~\ref{sec:profiling-procedure} must propagate. To exercise both at once, the pseudo-data are generated at a non-nominal nuisance configuration, $(\nu_{\text{shift}}, \nu_{\text{squeeze}}) = (0.5, -0.5)$, so that the fit must recover the distortion through $T_{\phi}$ while profiling the nuisances away from zero. All results below are evaluated on samples of $10^5$ events.

\subsection{Likelihood model and training}
The fit maximizes the extended joint likelihood over the kinematic and score features $(x, y)$ (Section~\ref{sec:methodology}), optimizing the Distribution of Interest (DoI) $T_{\phi}$ together with the nuisances $\nu$ so that the transformed reference reproduces the distorted pseudo-data. The per-flavour input densities $p_f(x \mid \nu)$ and $p_f(y \mid x, \nu)$ are pre-trained on the nominal simulation and frozen; their nuisance dependence is carried by the input systematic flow $T_\chi$ (Section~\ref{sec:factorizable_flows}), and the nuisance response of the measurement by the DoI systematic flow $T_{\psi}$ (Section~\ref{sec:doi_systematics}).

The nominal transformation $T_{\phi}^{\hat{\nu}}$ is realized as a stack of rational quadratic spline couplings~\cite{durkan2019neuralsplineflows}, with affine coupling layers~\cite{Kobyzev:2020nf,MADE} as a lighter alternative, and the residual $T_{\psi}(y \mid x, \nu)$ uses the factorizable construction of Ref.~\cite{Valsecchi:2026fnf}. Both are conditioned on the kinematics $x$ and on the event flavour, the latter encoded as a one-hot vector concatenated to $x$ before the conditioning networks, so that the correction is process-specific. In data the flavour of an event is unobserved, so the density entering the likelihood is the flavour mixture of Eq.~\ref{eq:mixture_model}: the transformation is evaluated for each class and the resulting per-class densities are summed with their mixture weights. The detailed architectures and hyperparameters are listed in Appendix~\ref{sec:appendix_hyperparameters}.

The model is trained following the two-step procedure of Section~\ref{sec:profiling-procedure}. In Step 1 this joint optimization is carried out with the AdamW optimizer~\cite{AdamW} and gradients from automatic differentiation~\cite{Paszke:2019pytorch}, giving the global best fit $\hat{T}_{\phi}^{\hat{\nu}}$ and $\hat{\nu}$. As discussed in Section~\ref{sec:training_dynamics}, the nuisances are assigned a larger learning rate than the high-dimensional transformation to balance their convergence; the likelihood scans obtained after this step are shown in Section~\ref{sec:results:likelihood_scans}. To propagate the statistical uncertainty of the measurement (Section~\ref{sec:stat_uncertainty}), Step 1 is repeated $K$ times as a Poisson bootstrap of the dataset, producing the ensemble $\{T_{\phi}^{(b)}\}$ of nominal transformations.

In Step 2 the nominal transformation is frozen and the residual systematic flow $T_{\psi}$ is trained, shared across the bootstrap ensemble, to capture how the DoI must deform across the nuisance space. The training is amortized (Section~\ref{sec:amortized}): rather than re-fitting at each nuisance point, $T_{\psi}$ is conditioned on $\nu$ and optimized over a fixed interval covering the range scanned at inference, with the per-event nuisance average evaluated on a Gauss-Legendre quadrature grid. A single run thus determines the systematic response over the whole nuisance space; the resulting scans are presented in Section~\ref{sec:results:step2}.

\subsection{Results}
\subsubsection{Step 1 - Global best-fit}
Step 1 jointly fits the transformation and the nuisances, and is repeated over the Poisson-bootstrap ensemble $\{T_{\phi}^{(b)}\}$ of Section~\ref{sec:stat_uncertainty}. We first determine the common anchor $\hat{\nu}$ from the likelihood scan, then validate the fit at that anchor in the observable space, and finally visualize the learned transformation.

\begin{figure}[!htbp]
    \centering
     \includegraphics[width=0.54\textwidth]{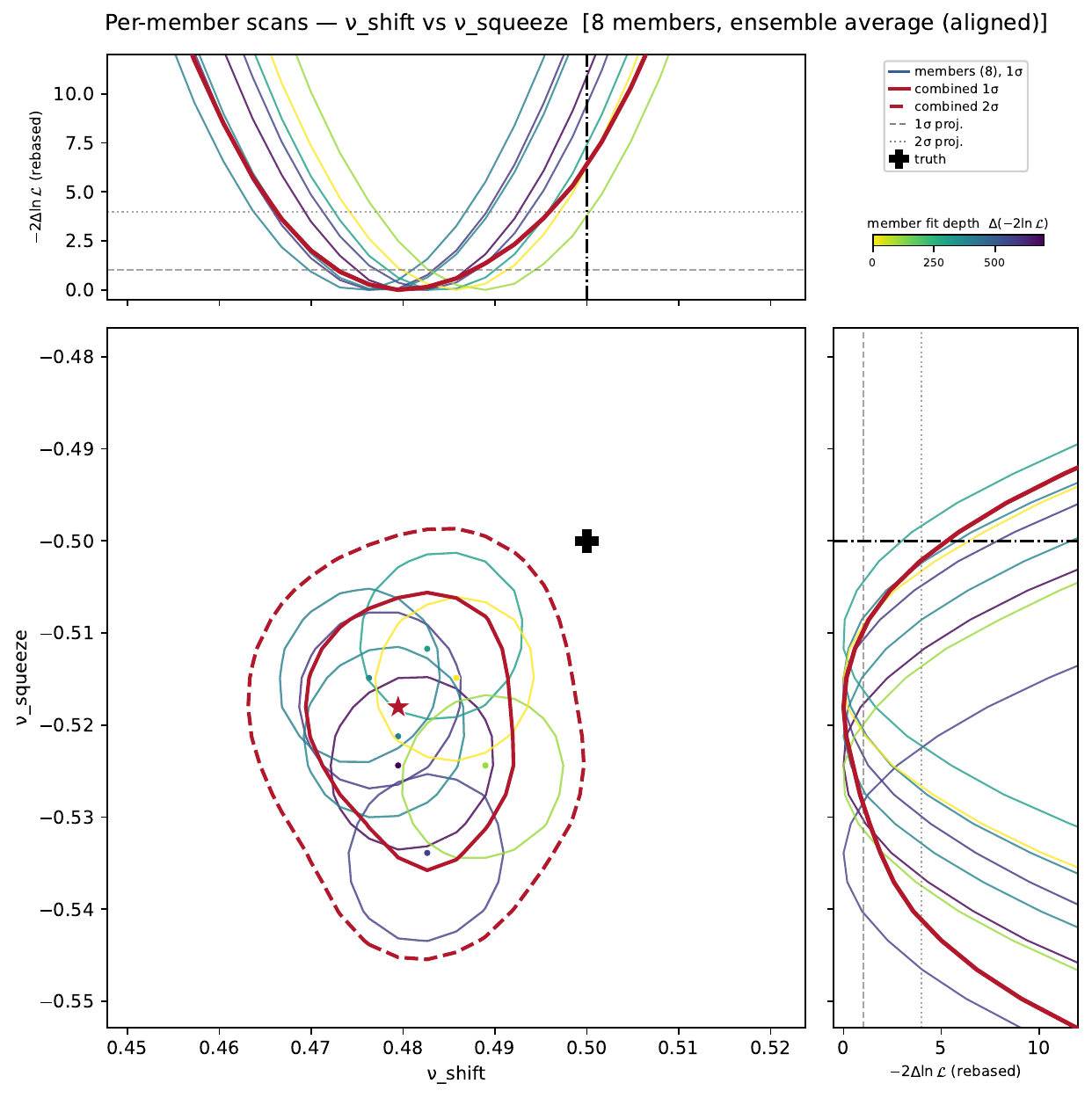}
    \includegraphics[width=0.45\textwidth]{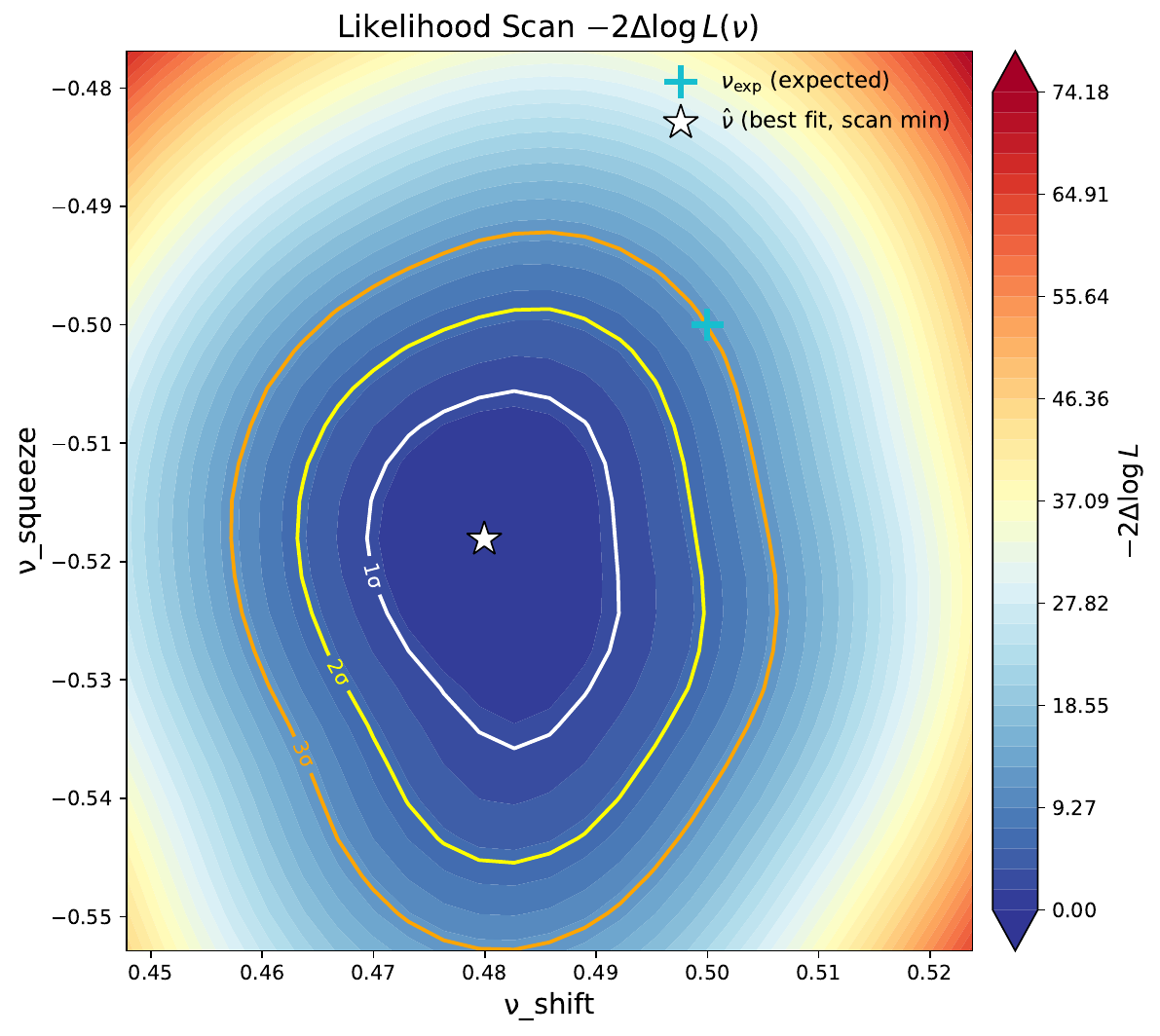}

    \caption{Step 1 likelihood scan over the two nuisances, combined across the Poisson-bootstrap ensemble by the ensemble-averaged (bagged) likelihood (Eq.~\ref{eq:bma}) to fold in the statistical uncertainty of the Distribution of Interest. \emph{Left:} the per-member two-dimensional scans (thin contours, coloured by fit depth) and their averaged combination (thick red, $1\sigma$ and $2\sigma$), with the marginal profiles along each nuisance; the per-member optima, the combined anchor $\hat{\nu}$, and the injected truth are overlaid. \emph{Right:} the combined profile $-2\ln L_{\mathrm{ens}}(\nu)$ as a likelihood surface with its $1\sigma$ and $2\sigma$ contours, the best-fit anchor $\hat{\nu}$ (star) and the injected value $\nu_{\mathrm{exp}}$ (square). The combined region is broadened by the member-to-member spread relative to any single member.}
    \label{fig:likelihood_scans}
\end{figure}

\paragraph{Likelihood scan and anchor}
\label{sec:results:likelihood_scans}
The Step 1 scan serves two purposes: it fixes the common anchor $\hat{\nu}$ around which the residual flow is expanded in Step 2, and it folds the statistical uncertainty of the Distribution of Interest into the inference. We scan the extended likelihood over the two nuisances with the transformation held fixed, and combine the Poisson-bootstrap ensemble: each member $T_{\phi}^{(b)}$ defines its own profile $L_b(\nu)$, and rather than selecting one we average them through the ensemble-averaged (bagged) likelihood of Eq.~\ref{eq:bma}, after rebasing each member to its own optimum and the combined profile to its own minimum (Eq.~\ref{eq:bma_rebased}). The minimum of the combined profile defines the anchor $\hat{\nu}$, while the averaging broadens the profile by the member-to-member spread, so that the resulting confidence region already includes the data-statistical uncertainty on the measurement on top of the nuisance constraints. 
Figure~\ref{fig:likelihood_scans} shows the scan in the plane of the two nuisances. The per-member scans cluster around a common minimum close to the injected truth, and their ensemble-averaged combination is wider than any single member, its $1\sigma$ and $2\sigma$ contours defining the anchor $\hat{\nu}$ together with its uncertainty.

The anchor falls close to the injected truth but not exactly on it, and this small residual offset is expected rather than a concern. The role of Step 1 is only to fix the anchor $\hat{\nu}$ around which the residual flow is expanded, not to deliver the final estimate of the nuisances, and its scan is evaluated with the transformation held fixed, which limits its sensitivity to the nuisances. The genuine sensitivity to the nuisances, and the final measurement, come from the Step 2 profiling described below, in which the conditional residual $T_{\psi}(\nu)$ lets the measured transformation deform with the nuisances.

\begin{figure}[!htbp]
    \centering
    \includegraphics[width=\textwidth]{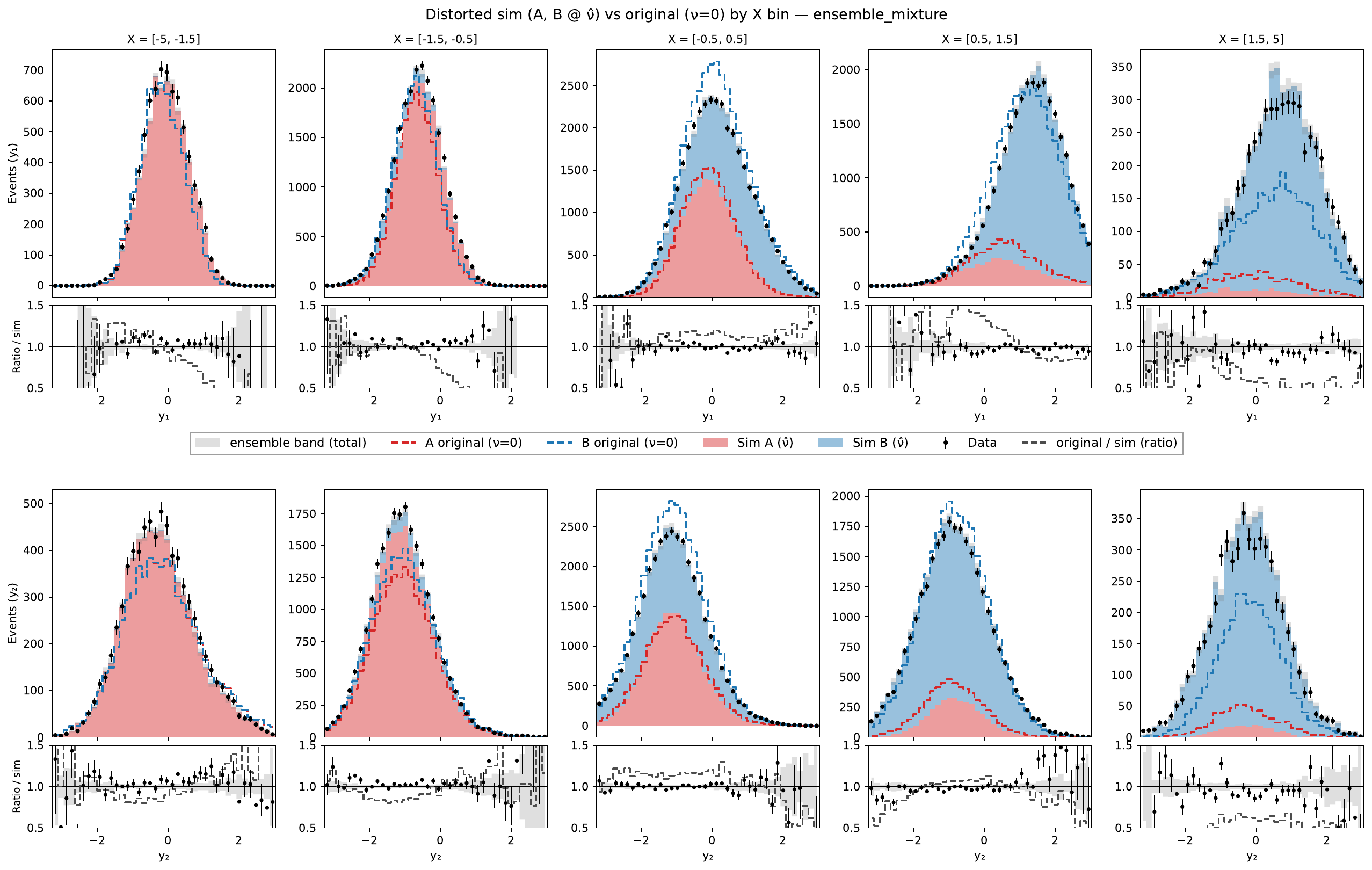}
    \caption{Step 1 postfit distributions in bins of $x$ (columns) for the score components $y_1$ (top) and $y_2$ (bottom). The prediction (sim), the reference sampled at $\hat{\nu}$ and transformed by the learned $T_{\phi}$, is compared with the distorted data (black points) and the undistorted expectation at $\nu = 0$ (dashed). The grey band is the spread over the Poisson-bootstrap ensemble $\{T_{\phi}^{(b)}\}$, the data-statistical uncertainty on the measurement (Section~\ref{sec:stat_uncertainty}). Lower panels: ratios to the data.}
    \label{fig:fit_quality}
\end{figure}

\paragraph{Postfit distributions}
Having fixed the anchor, we check the fit directly in the observable space. We sample the reference mixture of Eq.~\ref{eq:mixture_model} at the best-fit nuisances $\hat{\nu}$, apply the learned transformation $T_{\phi}$, and compare the prediction with the distorted data. Figure~\ref{fig:fit_quality} shows the resulting total prediction at the anchor $\hat{\nu}$ (histograms), the distorted data (black dots), together with the total undistorted expectation at $\nu = 0$ (dashed line), in bins of the kinematic variable $x$ for the two event classes. The prediction reproduces the distorted data across all kinematic bins, as confirmed by the ratios in the lower panels, indicating that the learned transformation captures the overall distortion in the feature space.

The grey band on the prediction quantifies its statistical uncertainty. As described in Section~\ref{sec:stat_uncertainty}, the band is the spread of the total prediction across the bootstrap ensemble $\{T_{\phi}^{(b)}\}$, all members being evaluated at $\hat{\nu}$ under common random draws so that it reflects the member-to-member variation of the transformation rather than the Monte Carlo sampling noise. It is the data-statistical uncertainty on the measured Distribution of Interest, shown directly in the observable space.

\begin{figure}[!htbp]
    \centering
    \includegraphics[width=\textwidth]{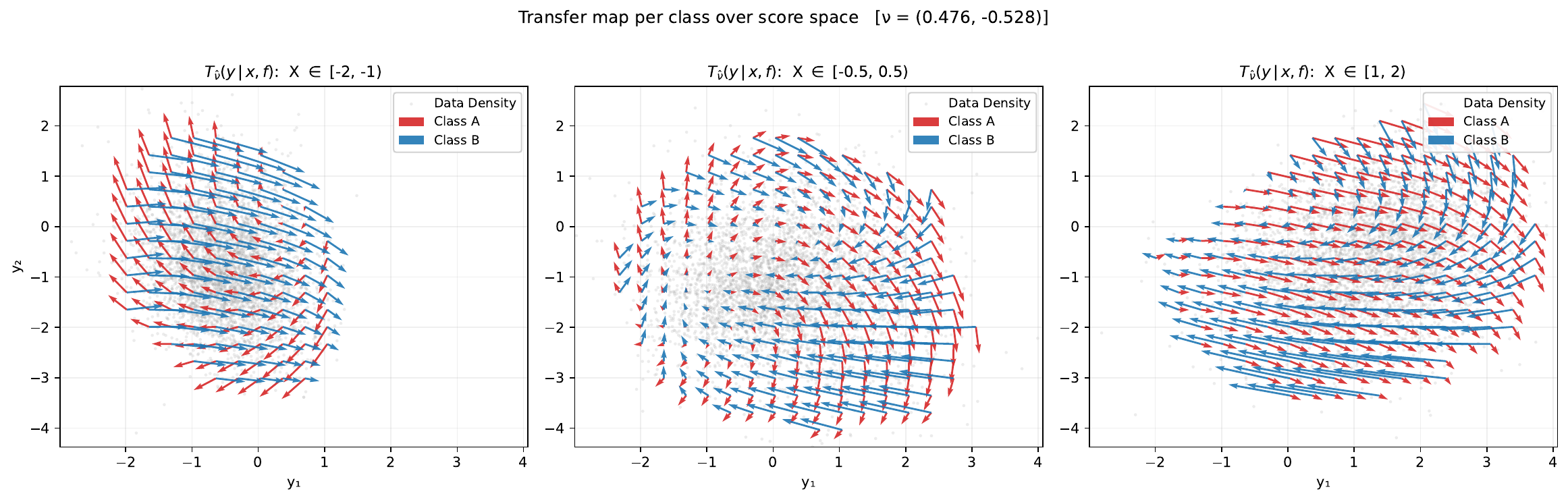}
    \caption{Learned Step 1 transformation $T_{\phi}^{(b)}$ for one representative member of the bootstrap ensemble, shown as a per-class displacement field over the score space $(y_1, y_2)$ in three slices of the kinematic variable $x$ (class A red, class B blue), evaluated at the best-fit anchor $\hat{\nu}$. The map reproduces the class-dependent data-simulation distortion injected into the pseudo-data; the remaining ensemble members differ only by the statistical spread of Figure~\ref{fig:fit_quality}.}
    \label{fig:transformation_step1}
\end{figure}

\paragraph{Learned transformation}
Finally, Figure~\ref{fig:transformation_step1} visualizes the learned transformation itself, for one representative member of the ensemble, as a per-class displacement field over the score space in three slices of the kinematic variable $x$. The map recovers the class-dependent distortion injected into the pseudo-data, so that the measured Distribution of Interest reproduces the data-simulation deformation; the other members of the ensemble differ from it only by the statistical spread quantified above.

\subsubsection{Step 2 - Amortized systematic-aware training}
\label{sec:results:step2}
With the anchor fixed, Step 2 trains the residual systematic flow $T_{\psi}$ that carries the nuisance dependence of the Distribution of Interest, composed with the frozen nominal map as in Eq.~\ref{eq:Texp}. A single $T_{\psi}$ is shared across the whole bootstrap ensemble: at each training step a member $T_{\phi}^{(b)}$ is drawn at random and composed with $T_{\psi}$, so that the residual learns the ensemble-averaged systematic response around the common anchor rather than being refit for each member (Section~\ref{sec:stat_uncertainty}). The training is amortized over the nuisance space (Section~\ref{sec:amortized}): $T_{\psi}$ is conditioned on $\nu$ and the likelihood is integrated over a Gauss-Legendre grid spanning the scanned interval, so that a single run learns the mapping from the nuisances to the DoI deformation across the entire space.

With the residual $T_{\psi}$ trained, we read off the systematic uncertainty by repeating the likelihood scan over the nuisances. The crucial difference from Step 1 is that the transformation is no longer frozen: \textbf{at each $\nu$ scan point the Distribution of Interest re-optimizes through the residual}, taking the value $T_{\phi}^{\hat{\nu}} \circ T_{\psi}(\nu)$. The scan therefore traces, point by point, how the measured distribution must deform to stay consistent with the data as the nuisances move away from the anchor, which is what makes the construction a genuine \textbf{profiling of the functional target}.

A single scan then delivers two uncertainties at once. Read along the nuisance axes, it is an ordinary profile likelihood and gives the confidence region on $\nu$, exactly as in a classical profiled fit. But each point on that region's contour also fixes a specific transformation $T_{\phi}^{\hat{\nu}} \circ T_{\psi}(\nu)$; as $\nu$ runs over the contour these transformations sweep out a family, and \textbf{their envelope is the systematic uncertainty on the Distribution of Interest itself}. The same scan thus measures both the nuisances and the distribution they deform: the first is read on the nuisance plane, the second in the space of transformations it induces.

\begin{figure}[!htbp]
    \centering
    \includegraphics[width=0.54\textwidth]{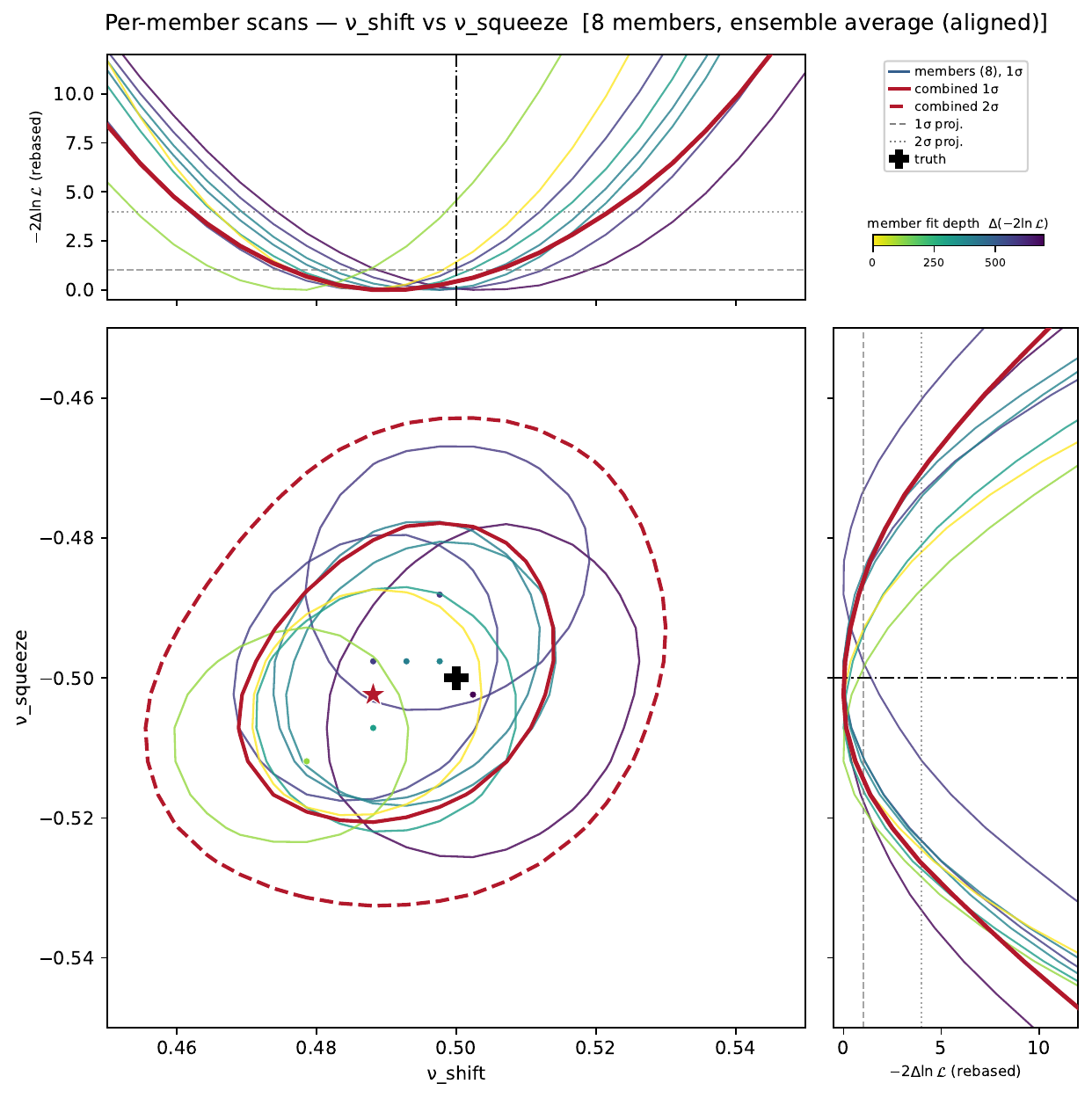}
    \includegraphics[width=0.45\textwidth]{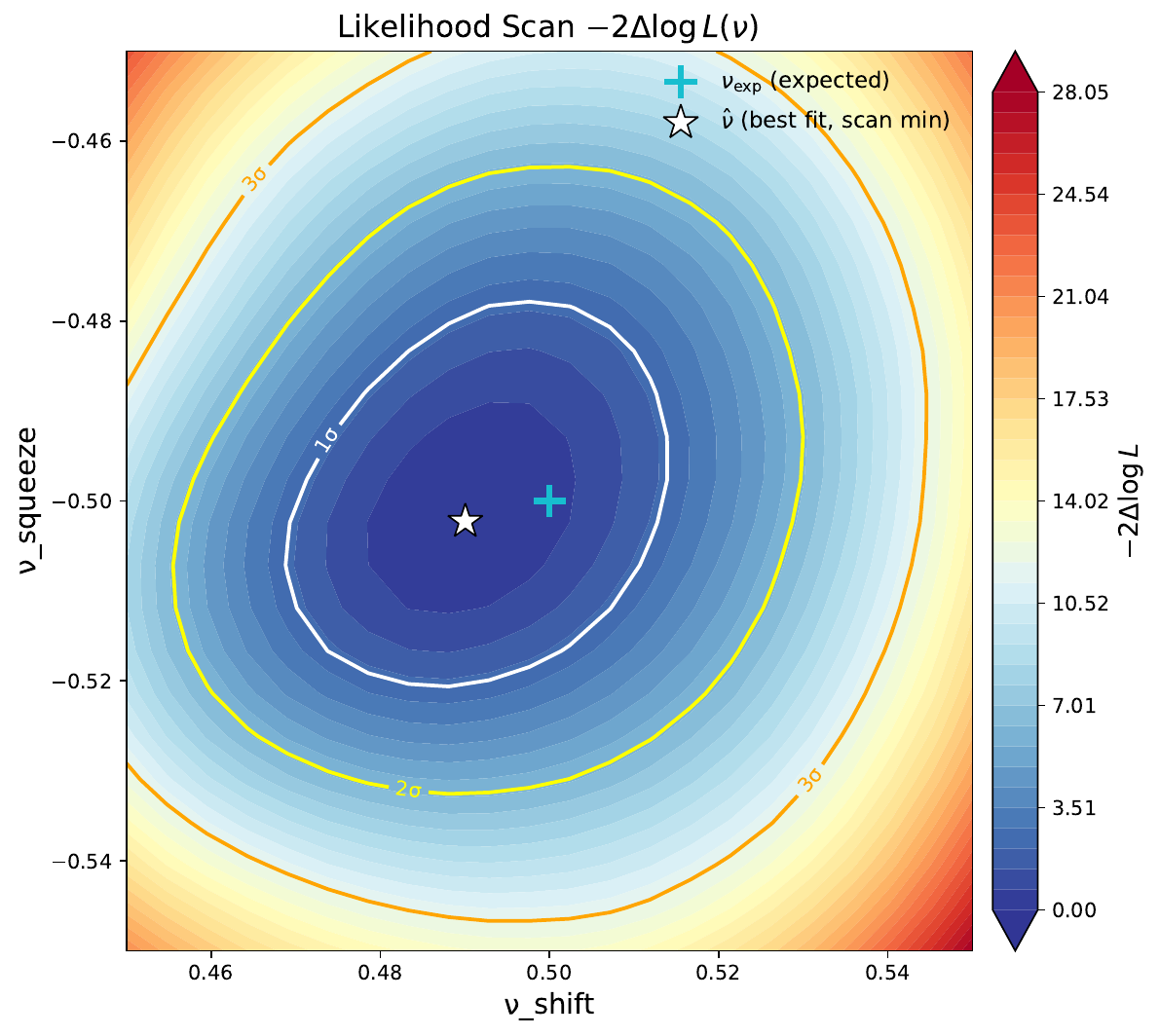}
    \caption{Likelihood scan after Step 2, with the full $\nu$-dependent Distribution of Interest $T_{\phi}^{\hat{\nu}} \circ T_{\psi}(\nu)$ re-evaluated at each scan point and combined across the Poisson-bootstrap ensemble by the ensemble-averaged (bagged) likelihood (Eq.~\ref{eq:bma}). \emph{Left:} the per-member scans (thin contours) and their averaged combination (thick red, $1\sigma$ and $2\sigma$), with the marginal profiles and the injected truth. \emph{Right:} the combined profile as a likelihood surface with its $1\sigma$ and $2\sigma$ contours. Relative to the Step 1 scan of Figure~\ref{fig:likelihood_scans}, the region is broadened by the systematic uncertainty now carried by the residual $T_{\psi}$.}
    \label{fig:likelihood_scans_step2}
\end{figure}

\begin{figure}[!htbp]
    \centering
    \includegraphics[width=0.6\textwidth]{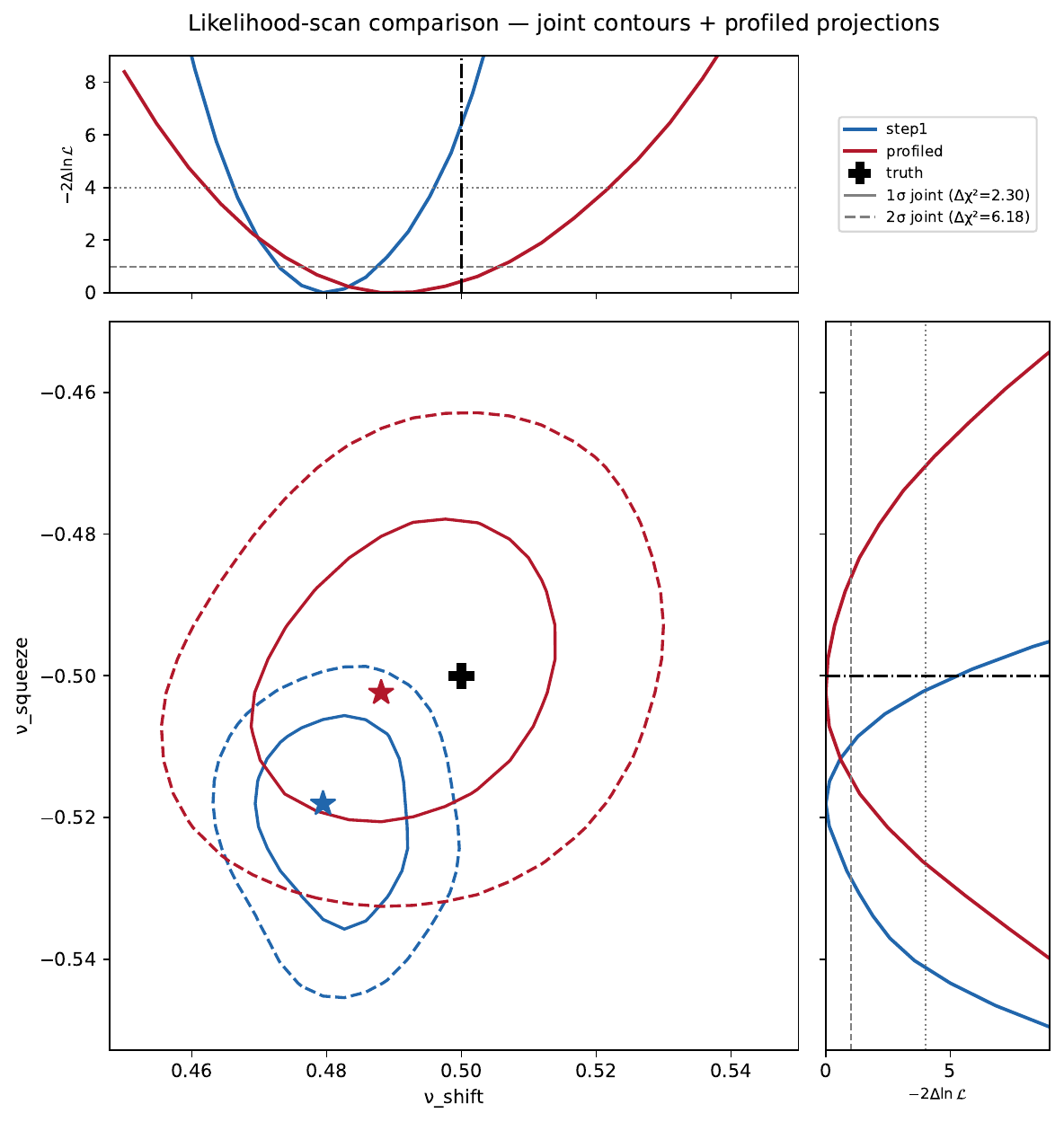}
    \caption{Likelihood scan before (Step 1, blue) and after (Step 2, red) the systematic-aware training, in the plane of the two nuisances with the marginal profiles. The solid and dashed contours are the $1\sigma$ and $2\sigma$ joint regions ($\Delta\chi^2 = 2.30$ and $6.18$); the stars mark the best fits and the cross the injected truth. Profiling broadens the region and moves the best fit toward the truth, the additional width being the systematic uncertainty absorbed by the Distribution of Interest.}
    \label{fig:scan_comparison}
\end{figure}

Averaging the scan across the Poisson-bootstrap ensemble, as in Step 1, folds the statistical uncertainty on top of the systematic one, so the final band on the Distribution of Interest carries both components. Figure~\ref{fig:likelihood_scans_step2} shows the combined Step 2 scan, and Figure~\ref{fig:scan_comparison} overlays it on the Step 1 result: the profiled region is broader, by the systematic uncertainty now absorbed into the DoI, and its best fit lies closer to the injected truth.

\begin{figure}[!htb]
    \centering
    \includegraphics[width=\textwidth]{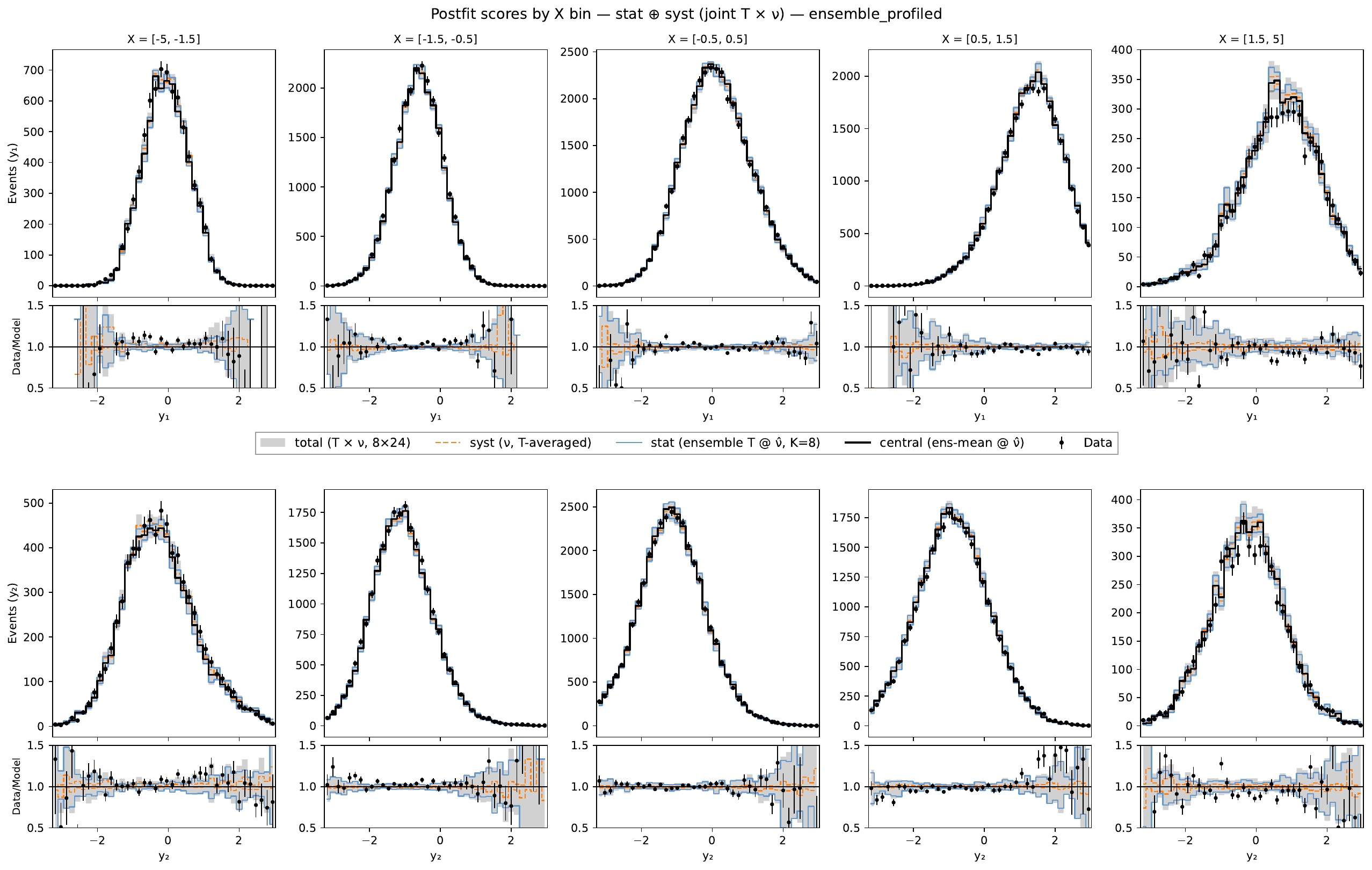}
    \caption{Post-fit score distributions ($y_1$, $y_2$ in rows) of the profiled DoI versus data, in bins of the kinematic observable $x_1$ (columns), each with a data/model ratio sub-panel. The central prediction (black) is the ensemble mean at the best fit $\hat{\nu}$; the statistical band is the per-bin standard deviation across the bootstrap ensemble $\{T_{\phi}^{(b)}\}$ sampled at $\hat{\nu}$ under common random numbers (only $T_{\phi}^{(b)}$ varies), and the systematic band is the prediction enveloped as $\nu$ runs over the $-2\Delta\ln L = 2.30$ contour (68\% CL, 2 d.o.f.) with the DoI averaged over the ensemble. The total band (filled) is the joint per-bin envelope over the full $K \times n_\nu$ grid of members and contour points, capturing the $(T_{\phi}, \nu)$ correlation exactly rather than by quadrature.}
    \label{fig:postfit_profiled_scan}
\end{figure}

Figure~\ref{fig:postfit_profiled_scan} shows the post-fit score distributions of the profiled DoI compared with the data. The two score components $y_1$ and $y_2$ are shown in rows and the events are partitioned into bins of the kinematic observable $x_1$ in columns, each panel carrying a lower sub-panel with the data-to-model ratio. The central prediction (solid black) is the ensemble-mean prediction at the best-fit nuisances $\hat{\nu}$, obtained by Monte Carlo sampling of the model and normalised to the observed yield in each bin. Two sources of uncertainty are propagated to the prediction and shown as nested bands. The statistical band is the data-statistical uncertainty of Section~\ref{sec:stat_uncertainty}, carried by the Poisson-bootstrap ensemble $\{T_{\phi}^{(b)}\}$ with each member composed with the frozen input densities and the shared residual $T_{\psi}$ at $\hat{\nu}$. All members are sampled under common random numbers, with identical flavour assignments, kinematics, and base scores, so that only $T_{\phi}^{(b)}$ differs between them; the band is the per-bin standard deviation across the members, and the common-random-number scheme ensures that it measures the genuine member-to-member variation rather than the Monte Carlo noise.

The systematic band measures how much the prediction shifts as the nuisances vary over the range allowed by the data. We take the $68\%$ confidence region of the Step 2 scan, the $-2\Delta\ln L = 2.30$ contour around $\hat{\nu}$ (two degrees of freedom), and in each bin record the spread of the prediction as $\nu$ runs along this contour, with the DoI averaged over the ensemble.

The two effects are then combined into a total band. Adding the statistical and systematic uncertainties in quadrature would assume them independent; instead we vary both at once: every bootstrap member is evaluated at every contour point, and the band is the per-bin envelope of this full $K \times n_\nu$ grid. Enumerating the joint $(T_{\phi}, \nu)$ variation this way captures any correlation between the two effects exactly, with no quadrature assumption. In each panel the statistical and systematic bands are drawn as edges over the filled total band, so their relative sizes can be read off directly.

\subsubsection{Uncertainty quantification and orthogonal decomposition}
With several correlated nuisances it is useful to identify the directions that actually drive the systematic uncertainty on the measurement. We apply the orthogonal decomposition of Section~\ref{sec:orthogonal_decomposition}: the Hessian of the negative log-likelihood with respect to $\nu$ is computed at the best fit $\hat{\nu}$ by automatic differentiation, and its eigen-decomposition yields orthogonal modes of the nuisance space, the eigenvalues measuring their constraining power and the eigenvectors the combinations of nuisances along which the Distribution of Interest varies independently.

\begin{figure}[!htbp]
\centering
\begin{subfigure}{\textwidth}
    \centering
    \includegraphics[width=0.62\textwidth]{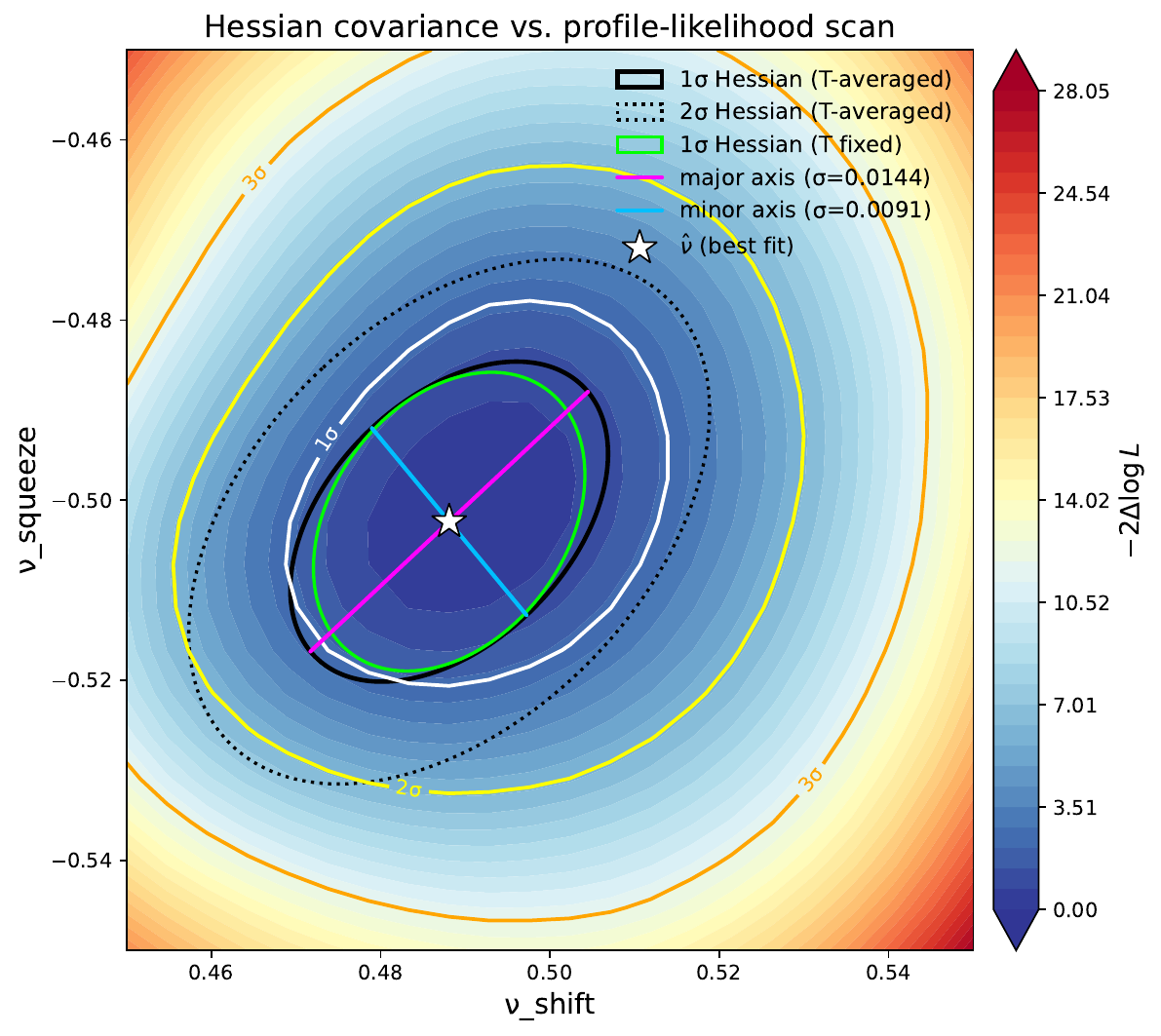}
    \caption{Hessian covariance at $\hat{\nu}$ over the profile-likelihood scan: the $1\sigma$ and $2\sigma$ ensemble-averaged contours (white, statistical and systematic), the $1\sigma$ fixed-transformation contour (green, systematic only), and the major and minor eigenvectors. The star marks $\hat{\nu}$.}
    \label{fig:ortho_hessian}
\end{subfigure}

\begin{subfigure}{\textwidth}
    \centering
    \includegraphics[width=0.8\textwidth]{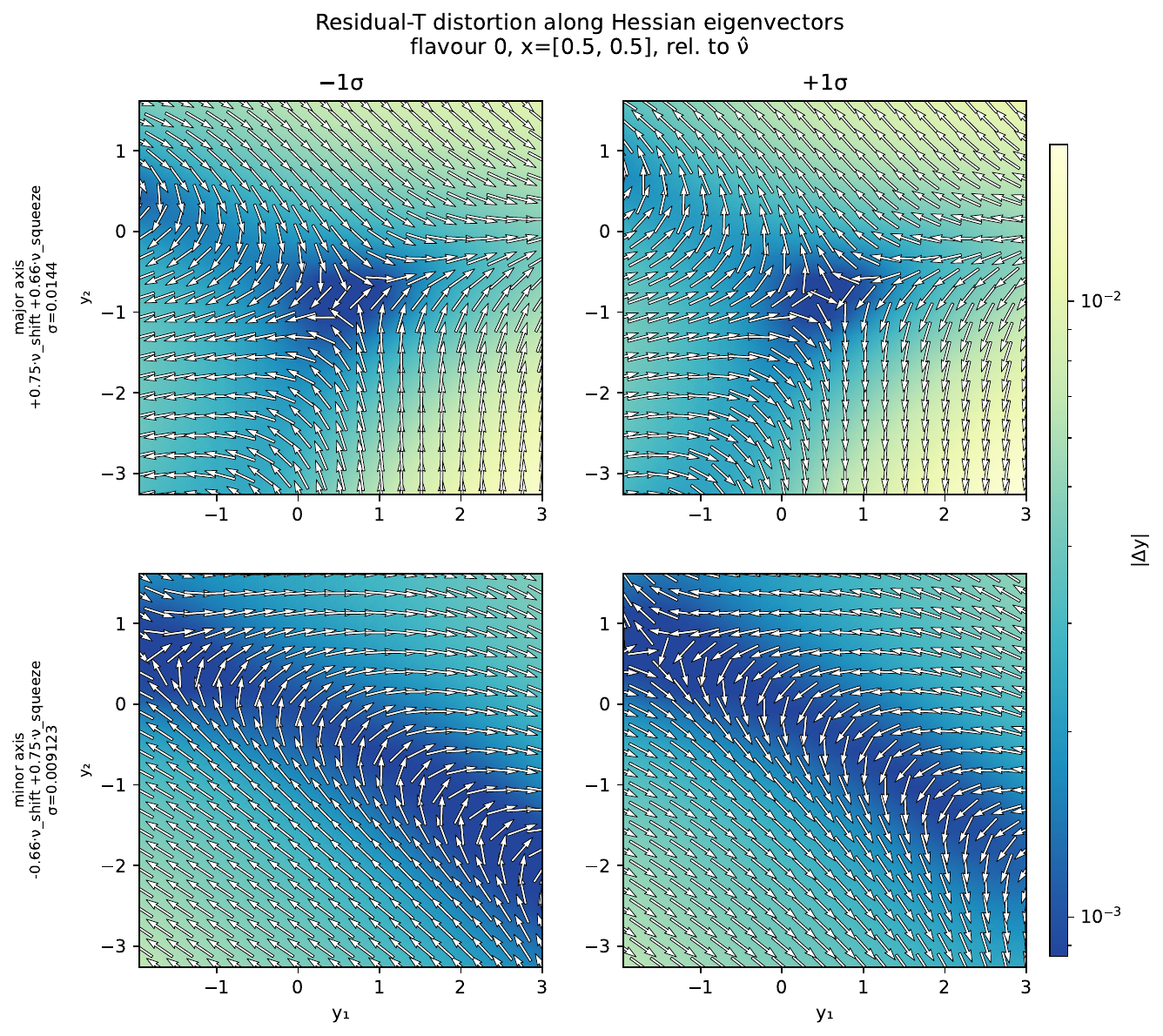}
    \caption{Score displacement induced by the residual $T_{\psi}$ at $\pm 1\sigma$ (columns) along the major (top) and minor (bottom) eigenvectors, for class A at $x = (0.5, 0.5)$; the colour gives $|\Delta y|$.}
    \label{fig:ortho_pull}
\end{subfigure}
\caption{Orthogonal decomposition of the systematic uncertainty after Step 2 (Section~\ref{sec:orthogonal_decomposition}).}
\label{fig:orthogonal_decomposition}
\end{figure}

Figure~\ref{fig:ortho_hessian} overlays the resulting local-Gaussian covariance on the profile-likelihood scan, with the two eigenvectors drawn as the major and minor axes. Including the bootstrap ensemble widens the covariance: the ellipse of the ensemble-averaged likelihood is larger than the one obtained at fixed transformation, the difference being the statistical contribution of Section~\ref{sec:stat_uncertainty}, derived as a local covariance in Appendix~\ref{sec:appendix_bma_hessian}. The Gaussian approximation is only a local description here, since the profiled likelihood is not exactly quadratic, but it still isolates the principal modes of variation cleanly.

Figure~\ref{fig:ortho_pull} then visualizes the principal systematic variations themselves. Evaluating the residual $T_{\psi}$ at $\pm 1\sigma$ along each eigenvector and taking its displacement of the score relative to the best fit reveals the deformation that each mode induces on the measurement, decoupled from the original correlated parametrization. The two eigenvectors act on the score in distinct, orthogonal ways, making explicit the combined shifts of the nuisances to which the measured transformation is most and least sensitive.

\subsubsection{Empirical validation of the estimator coverage and bias}
\label{sec:results:coverage}

To validate the statistical properties of the extracted uncertainties, we perform an empirical coverage test on ensembles of statistically independent toy datasets, each generated at the same non-nominal truth $(\nu_{\text{shift}}, \nu_{\text{squeeze}}) = (0.5, -0.5)$ used throughout. For each toy the full two-step procedure is repeated, a global fit and Poisson bootstrap (Step 1) followed by amortized systematic profiling (Step 2), and the $68\%$ confidence region is tested for inclusion of the truth. We run the study in two configurations of the input systematic model that carries the nuisance dependence of the base kinematic and score densities (Section~\ref{sec:factorizable_flows}): the \emph{factorized} model used for the main results, whose joint nuisance response is recovered by summation over the per-nuisance terms, and a \emph{correlated} model in which the pairwise cross-term between the two nuisances, the bilinear $\nu_{\text{shift}}\,\nu_{\text{squeeze}}$ term of Eq.~\ref{eq:fnf_poly}, is additionally modelled (Appendix~\ref{sec:appendix_crossterms}). The factorized model is evaluated on 40 toys and the correlated one on 100.

\begin{figure}[!htbp]
    \centering
    \includegraphics[width=0.48\textwidth]{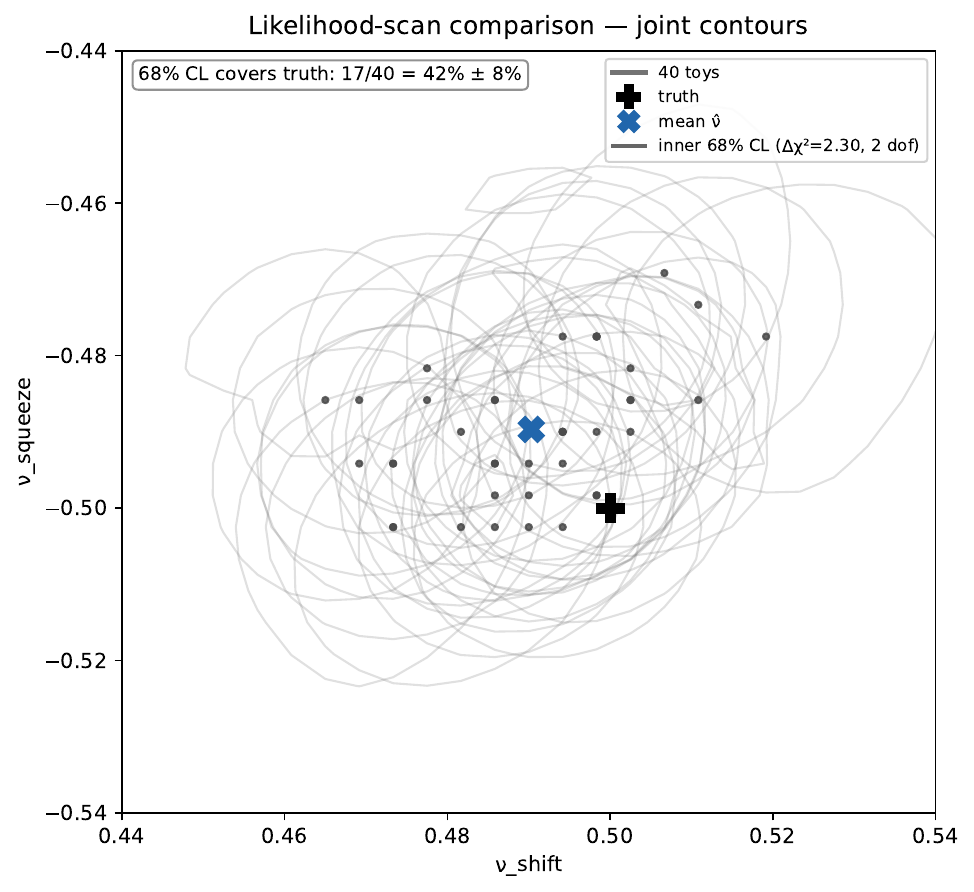}
    \includegraphics[width=0.48\textwidth]{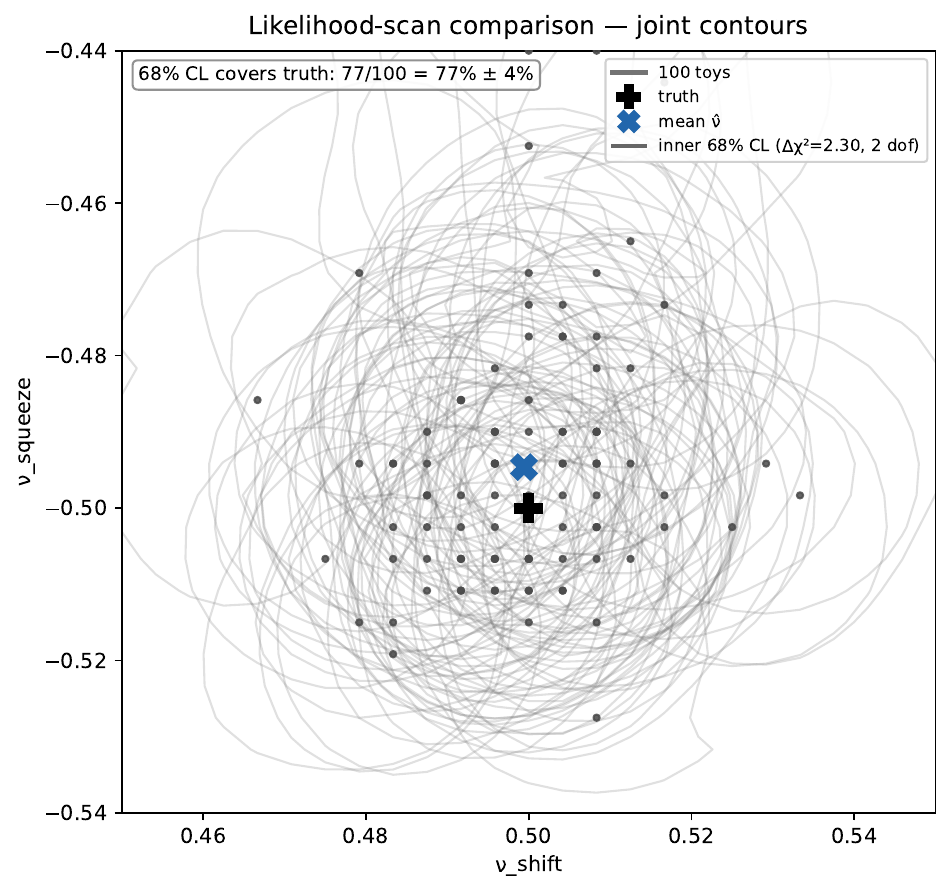}
    \caption{Empirical coverage test on statistically independent toy datasets generated at the injected truth $(\nu_{\text{shift}}, \nu_{\text{squeeze}}) = (0.5, -0.5)$ (black cross). Thin grey curves are the per-toy $68\%$ joint confidence regions ($\Delta\chi^2 = 2.30$, two degrees of freedom) from the full two-step fit; the blue cross marks the mean best fit $\bar{\nu}$ over the ensemble. \emph{Left:} factorized input systematic model, used for the main results, whose mean best fit is offset from the truth and which covers $17/40 = 42\%$. \emph{Right:} correlated model with the pairwise nuisance cross-terms enabled, whose mean best fit lands on the truth and which covers $77/100 = 77\%$ at the same nominal threshold.}
    \label{fig:coverage_test1}
\end{figure}

Figure~\ref{fig:coverage_test1} compares the two. 
With the factorized model (left panel), a naive evaluation of the joint $68\%$ coverage with respect to the true nuisances yields an inclusion rate of only $17/40 = 42\%$ at the standard Wilks threshold ($\Delta\chi^2 \leq 2.30$ for two degrees of freedom); reaching nominal coverage would require inflating the threshold to $\Delta\chi^2 \approx 4.71$.

Inspection of the ensemble shows that the under-coverage of the factorized model is dominated by a bias in the central point estimator, the global minimum $\hat{\nu}$ of Step 1, rather than by a miscalibration of the variance extracted in Step 2: the mean of the best-fit points across the 40 toys is offset from the truth by roughly three standard errors of the mean, while the per-toy contours have the right size. 
This is made explicit by a \emph{bias-corrected} coverage, the inclusion rate of the confidence regions with respect to the empirical mean $\bar{\nu}$ of the toy ensemble rather than the truth, which neutralizes the central offset. Evaluated this way the factorized model recovers nominal behaviour, $70\%$ at $1\sigma$ ($\Delta\chi^2 \leq 2.30$) and $95\%$ at $2\sigma$ ($\Delta\chi^2 \leq 6.18$), confirming that the size and shape of the contours produced by the amortized profiling are correctly calibrated and that the deficit is entirely a displacement of the central value.

The origin of this displacement is the fidelity of the systematic model at the injected point. The truth $(0.5, -0.5)$ drives both nuisances strongly and simultaneously, precisely the regime in which the factorized parameterization, which sums the per-nuisance responses and omits their bilinear $\nu_{\text{shift}}\,\nu_{\text{squeeze}}$ interaction, the cross-term $\phi$ of Eq.~\ref{eq:fnf_poly}, is least accurate (Ref.~\cite{Valsecchi:2026fnf}, Appendix~\ref{sec:appendix_crossterms}). The missing cross-term is a genuine correlated deformation of the densities that the factorized model cannot reproduce, and in the joint Step 1 optimization the fit absorbs it into a small compensating shift of $\hat{\nu}$, the bias seen in the left panel. The displacement is therefore a property of the systematic model, not of the profiling: Step 2 faithfully propagates whatever transformation it is given, so the correctly sized contours are simply centred on a slightly displaced point. 

Explicitly modelling this cross-term in the kinematic and score systematic flows removes the bias. The right panel of Figure~\ref{fig:coverage_test1} repeats the study with the correlated systematic model, in which the pairwise cross-terms of the base kinematic and score flows are enabled and calibrated on combined $\pm 1\sigma$ variations of the two nuisances. The mean best fit now lands on the truth and the coverage rises to $77\%$ at the standard Wilks threshold, with no recentering. The same Step 2 machinery that under-covered with the factorized model slightly overcovers once the systematic response it profiles is itself accurate. The cross-terms here are enabled only in the input systematic model that describes the data; the DoI residual $T_{\psi}$ that carries the nuisance dependence of the measurement is kept factorized, as in all results of this work, so the correction acts on the modelled data templates and not on the measured transformation itself (Appendix~\ref{sec:appendix_crossterms}).

Two conclusions follow. The residual bias of the factorized fit is not a limitation of the amortized profiling but a controlled systematic-modelling effect: when the nuisance response is modelled precisely, including its correlated cross-terms, the estimator is unbiased and the intervals cover close to the nominal rate. And this fidelity is available on demand within the same construction, since the cross-terms require combined-variation samples and a number of interaction terms that grows with the number of nuisance \emph{pairs}, a quadratic rather than exponential cost; they can be enabled selectively for the dominant correlated pairs while the bulk of the systematic budget is kept in the cheap, linearly-scaling factorized form. The factorized model used for the main results is thus the conservative default, and the coverage study shows both that its profiling is correctly calibrated and that its small central bias is removed by precise modelling of the systematic uncertainty.

A full characterization of the coverage of the proposed estimator chain is, however, left to future work. The ensembles used here are modest, 40 toys for the factorized model and 100 for the correlated one, and probe a single injected truth; a systematic study across a range of truth points spanning the nuisance space, with a varying number of nuisances and substantially larger ensembles, is needed to map the inclusion rate as a function of these factors and to reduce the binomial uncertainty on the quoted rates. In particular, the mild overcoverage of the correlated model, $77\%$ against the nominal $68\%$, deserves closer scrutiny. It may reflect a conservative propagation of the finite-sample uncertainty through the Poisson-bootstrap and ensemble-averaging chain, and disentangling such a residual effect from the limited toy statistics will require the larger ensembles mentioned above. Establishing the asymptotic coverage of the full bootstrap-plus-bagging construction, and the conditions under which it attains its nominal rate, is an important direction for further validation of the method.

\section{Discussion and Outlook}
\label{sec:discussion}
This work demonstrates a complete measurement of a Distribution of Interest: an invertible transformation that maps the reference model onto the data, delivered with its full uncertainty budget. The systematic part comes from profiling the nuisances with Factorizable Normalizing Flows (FNF). A single amortized training replaces the expensive profiling scan, at a cost that grows linearly rather than exponentially with the number of nuisances, a robust alternative to the classifier-interpolation and parametric-ansatz methods of Section~\ref{sec:related_work}. The statistical part is the finite-sample uncertainty of the fitted transformation; it is captured by the Poisson-bootstrap ensemble and propagated through the same likelihood scan by the ensemble-averaged (bagged) likelihood.

In practice, we profile the parameters of a neural network. The network is the transformation that defines the measurement, its parameters play the role of the parameter of interest, and the fit returns both the best-fit map and an uncertainty on the learned function, shown as the statistical and systematic bands of Section~\ref{sec:results:step2}. The two sources are combined in a single, self-consistent unbinned likelihood rather than added by hand after the fit, giving the functional measurement a complete uncertainty budget with frequentist meaning.

On the synthetic dataset the procedure behaves as a well-calibrated estimator: the global fit recovers the injected nuisance values (Figures~\ref{fig:likelihood_scans} and~\ref{fig:likelihood_scans_step2}) and the post-fit prediction reproduces the distorted data within the quoted bands, indicating that the measured transformation is an unbiased estimator of the underlying data-simulation distortion. The frequentist coverage study of Section~\ref{sec:results:coverage} corroborates this on ensembles of pseudo-experiments carried through the full bootstrap and ensemble-averaging chain, confirming that the confidence regions are correctly sized and, once the systematic response is modelled accurately, are slightly conservative.

Because the result is a reusable transformation rather than a set of fitted numbers, the same construction serves several measurement tasks in HEP:
\begin{itemize}
    \item \textbf{Detector calibration}: The DoI paradigm can be used to formulate simulation-to-data correction measurements, in which the fitted transformation is the calibration of the detector response measured directly from data. This approach allows to correct the simulation continuously over the full observable space, rather than through a per-bin scale factor or a single global constant, and it carries a statistical and systematic uncertainty on the correction itself (Section~\ref{sec:results:step2}) that downstream analyses can propagate to any quantity computed from the corrected simulation.
    \item \textbf{Continuous parameter estimation}: Targeting a transformation instead of a scalar generalizes parameter estimation from discrete or binned summaries to continuous quantities. A differential cross-section can be read off as a smooth function of the observables, evaluated at any point without committing to bin edges, while a continuous physics parameter such as a mass or a coupling is inferred directly from the unbinned likelihood, which retains the sensitivity carried by the tails and correlations that a binned summary averages away.
    \item \textbf{Unfolding}: Recovering the generator-level, particle-level spectrum from the reconstructed, detector-level distributions can be structured as a DoI measurement. Recent generative-unfolding results~\cite{Butter:2025dm,T2K:2025omnifold,Butter:2025analysis,Bahl_2025,butter2025simulationpriorindependentneuralunfolding} reach high precision in idealised settings but leave the incorporation of experimental systematics open, owing to the difficulty of building them into the underlying generative machine-learning models. Our strategy targets that gap, enabling systematic-aware unfolding with a complete uncertainty budget. A dedicated study applying the method to unfolding is in preparation.
    \item \textbf{Analysis preservation and reinterpretation}: A major challenge in modern HEP is the publication and preservation of full likelihoods for future reinterpretation. By encapsulating the complete unbinned likelihood, including the factorized systematic profiling, within a differentiable neural network, this framework circumvents the need for publishing massive, inflexible statistical workspaces. The resulting trained model acts as a standalone, reusable statistical representation of the data: it can be evaluated as a likelihood and, being a normalizing flow, sampled directly to generate pseudo-datasets, supporting toy-based statistical studies and reinterpretation without rerunning the original analysis.
\end{itemize}

\section{Conclusion}
\label{sec:conclusion}

We have presented a framework for unbinned likelihood fits that measures a distribution of interest while propagating both its systematic and its statistical uncertainty. Four ingredients combine into a single inference: (1) the \emph{Distribution of Interest}, an invertible transformation that generalizes the measurement from scalar parameters to a full distribution; (2) Factorizable Normalizing Flows, which decompose systematic effects into independent, interpretable contributions; (3) an amortized training that learns the response of the likelihood to the nuisances across their whole range in a single pass, replacing the repeated fits of a profiling scan with an upfront cost; and (4) a Poisson-bootstrap ensemble combined by an ensemble-averaged (bagged) likelihood, which propagates the finite-sample uncertainty of the learned transformation into the result.

We validated the method on a synthetic dataset with two systematic sources, where it reproduces the injected data-simulation distortion and delivers a combined statistical and systematic uncertainty on the measured transformation. The construction is modular and, by design, its cost scales linearly rather than exponentially with the number of nuisances.

By elevating the target of unbinned likelihood fits from scalar parameters to functional distributions, and equipping them with a tractable, complete statistical-plus-systematic uncertainty budget, this framework bridges the gap between modern generative machine learning and the rigorous statistical demands of High Energy Physics. It offers a paradigm shift in how differential measurements, detector calibrations, and unfolding are performed, paving the way for fully unbinned, high-dimensional precision physics at the High-Luminosity LHC and beyond.

\section*{Acknowledgements}

D.V. acknowledges the support from the Swiss National Science Foundation under contract number 10003769.

\section*{Data and code availability}

The software implementing the method and reproducing the results of this work is publicly available at~\cite{Valsecchi:2026code,Valsecchi:2026fnf_code}.

\begin{appendices}
\numberwithin{equation}{section}

\section{Details about synthetic dataset}
\label{sec:appendix_dataset}

The generation proceeds in three stages: sampling the kinematics $x$, drawing the score $y$ conditional on the kinematics, and, for the pseudo-data only, applying the data-simulation distortion map. The two nuisances act with opposite sign on the two classes; we write $s_f = -1$ for class $A$ and $s_f = +1$ for class $B$. The nuisance deformations follow the construction of Ref.~\cite{Valsecchi:2026fnf}, where the same dataset is used; we summarize them here and give the distortion map, which is specific to this work, in full.

\paragraph{Kinematics.}
The kinematic vector $x \in \mathbb{R}^2$ is drawn from a class-conditional Gaussian with diagonal covariance,
\begin{equation}
x \mid f \sim \mathcal{N}\!\big(\mu_f,\ \mathrm{diag}(\sigma_f^2)\big), \qquad
\mu_A = (-0.5,\, 0),\ \ \sigma_A = (0.9,\, 0.6), \qquad
\mu_B = (+0.5,\, 0),\ \ \sigma_B = (0.6,\, 0.4).
\end{equation}
The squeeze nuisance applies a volume-preserving, axis-anti-correlated scaling about each centroid, and the shift nuisance translates the centroids along $\hat{d} = (1, 0)$,
\begin{equation}
x \to \mu_f + \mathrm{diag}\!\big(e^{+a_f}, e^{-a_f}\big)\,(x - \mu_f) + s_f\, \nu_{\text{shift}}\, s_{\text{shift}}\, \hat{d},
\qquad a_f = -s_f\, \nu_{\text{squeeze}}\, s_{\text{squeeze}},
\end{equation}
with $s_{\text{shift}} = 0.3$ and $s_{\text{squeeze}} = 0.2$. The squeeze preserves the cluster area ($\det = 1$): class $A$ stretches the first kinematic axis and compresses the second, class $B$ the reverse. Figure~\ref{fig:nuisance_xy} shows the effect of the nuisances on $p_f(x)$.

\paragraph{Score features.}
The score $y \in \mathbb{R}^2$ is drawn from a conditional bivariate Gaussian $p_f(y \mid x, \nu) = \mathcal{N}(\mu, \Sigma)$. The mean carries the nominal kinematic dependence plus a linear, $x$-dependent shift driven by $\nu_{\text{shift}}$,
\begin{align}
\mu_1 &= \sin(1.5\,x_1) + 0.3\,x_2 + s_f\, \nu_{\text{shift}}\, c_y \tanh(x_1), \\
\mu_2 &= 0.3\,x_1^2 - 1.2 + 0.5 \sin(x_2) - s_f\, \nu_{\text{shift}}\, c_y \tanh(x_1),
\end{align}
and the covariance is built from per-axis spreads, rescaled by $\nu_{\text{squeeze}}$, together with an $x$-dependent correlation,
\begin{align}
\sigma_1 &= \mathrm{softplus}(0.4\,x_1 + 0.1)\; e^{+s_f\, \nu_{\text{squeeze}}\, d_y}, &
\sigma_2 &= \mathrm{softplus}(-0.2\,x_1 + 0.4)\; e^{-s_f\, \nu_{\text{squeeze}}\, d_y}, \\
\rho &= 0.8 \tanh\!\big(0.5\,(x_1 + x_2)\big). & &
\end{align}
The score response scales are $c_y = 0.3$ ($\nu_{\text{shift}}$, a linear mean shift) and $d_y = 0.2$ ($\nu_{\text{squeeze}}$, a determinant-preserving rescaling of the spread). Figure~\ref{fig:nuisance_xy} shows their effect on $p_f(y \mid x)$.

\begin{figure}[!htb]
  \centering
  \includegraphics[width=\textwidth]{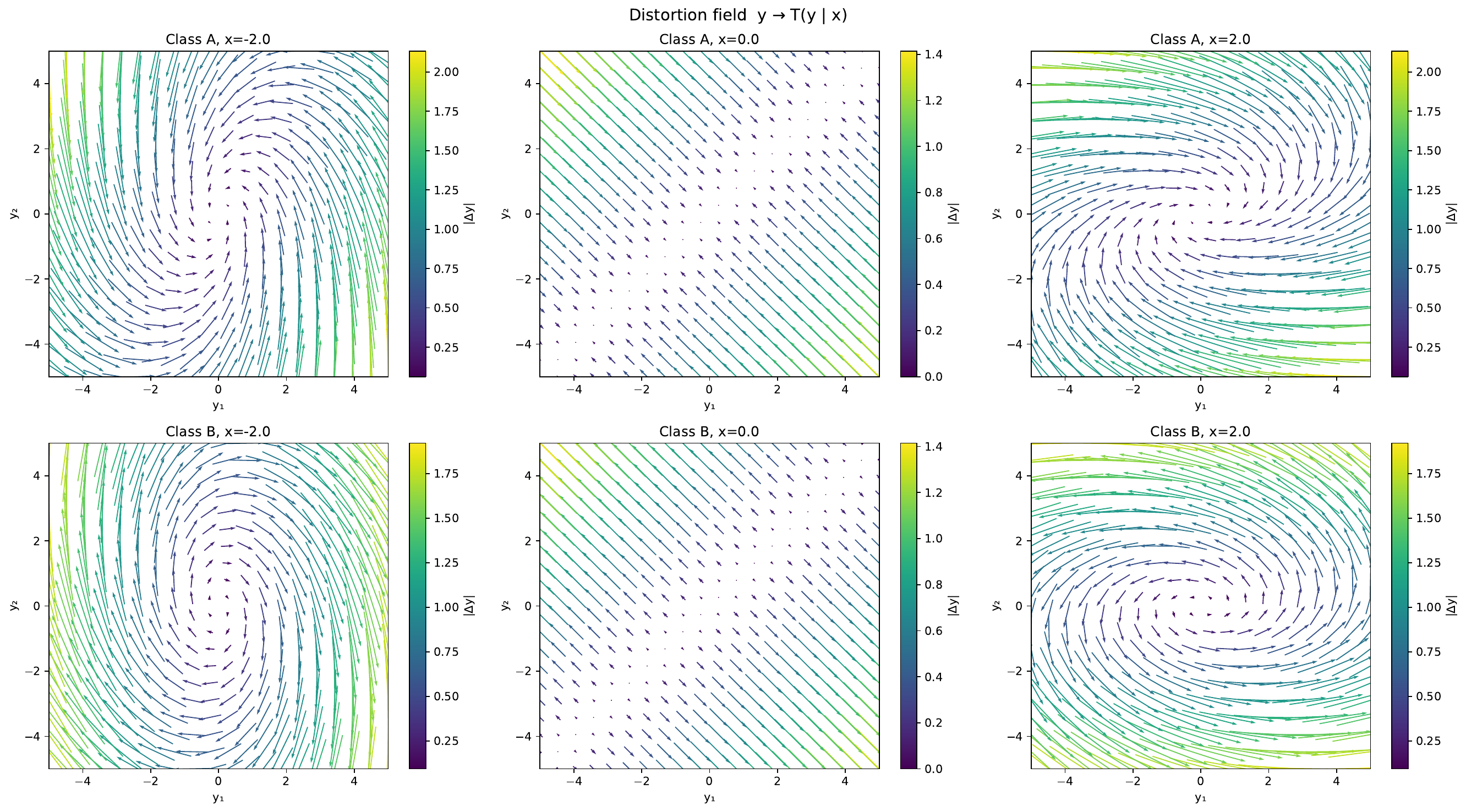}
  \caption{Effect of the distortion map on the score for different values of the kinematic variable $x$ (columns) and different classes (rows).}
  \label{fig:distorsion}
\end{figure}

\paragraph{Data-simulation distortion.}
The pseudo-data are obtained by applying a fixed distortion map to the score, standing in for the data-simulation mismodelling. A class-dependent rotation $R(\theta)$, whose angle decays with the kinematic radius $r = |x|$ and is antisymmetric in $x_1$, is followed by a small class-dependent shear,
\begin{equation}
\theta(x, f) = s_f\, \frac{2\beta}{0.7\,r + 0.5}\, \tanh\!\big(2\,x_1\big), \qquad y' = R(\theta)\, y,
\end{equation}
\begin{equation}
y_{\text{final}} = y' + s_f\, \kappa\,(y'_1 - y'_2)\begin{pmatrix} 1 \\ -1 \end{pmatrix},
\end{equation}
with distortion strength $\beta = 0.2$ and shear coefficient $\kappa = 0.1$. The map depends only on $(y, x, f)$ and carries no dependence on the nuisances, so it is precisely the data-simulation difference that the DoI transformation $T_{\phi}$ is meant to measure. Figure~\ref{fig:distorsion} shows its effect on the score for several values of $x$.

\section{Hyperparameters and Model Architectures}
\label{sec:appendix_hyperparameters}
All models are implemented in \texttt{PyTorch}~\cite{Paszke:2019pytorch}, with the Normalizing Flows built using the \texttt{Zuko} library~\cite{rozet2022zuko} and trained with the AdamW optimizer~\cite{AdamW}. The nominal input densities are modelled with Neural Spline Flows (NSF)~\cite{durkan2019neuralsplineflows}: the kinematic density $p(x \mid c)$ uses 2 transforms with 20 bins and hidden layers of $128 \times 3$ nodes, and the conditional score density $p(y \mid x, c)$ uses 3 transforms with 20 bins and hidden layers of $128 \times 3$ nodes. Both are trained for 100 epochs on samples drawn on the fly from the nominal generator, with a batch size of 1024 and a learning rate of $10^{-4}$.

The systematic dependence of the input densities is modelled with the Factorizable Normalizing Flows of Ref.~\cite{Valsecchi:2026fnf}, one per factor and conditioned on the two nuisances. The kinematic input flow $T_{\chi}^{x}$ uses 1 residual layer with hidden layers of $128 \times 2$ nodes, and the score input flow $T_{\chi}^{y}$ uses 2 residual layers with hidden layers of $128 \times 3$ nodes. They are calibrated on the per-nuisance $\pm 1\sigma$ samples for 60 epochs, with a batch size of 1024 and a learning rate of $10^{-4}$.

The nominal DoI transformation $T_{\phi}$ is an NSF with 2 transforms, 16 bins and hidden layers of $256 \times 2$ nodes, conditioned on the kinematics and the event flavour. In Step 1 it is fitted jointly with the two nuisances on $10^5$ events for 200 epochs (batch size 1024), using a learning rate of $10^{-4}$ for the transformation and a larger $4 \times 10^{-3}$ for the nuisances. The data-statistical uncertainty is estimated from a Poisson-bootstrap ensemble of $K = 8$ such fits.

The DoI systematic flow $T_{\psi}$ is a Factorizable residual transformation with 3 layers, hidden layers of $256 \times 2$ nodes, and 2 nuisance parameters. In Step 2 it is trained for 300 epochs (batch size 1024, learning rate $10^{-4}$), shared across the ensemble, with the nuisances sampled on a Gauss-Legendre grid of 6 nodes per nuisance over the interval $[-0.6, 0.6]$ in each direction.

\section{Learned residual response}
\label{sec:appendix_residual}

\begin{figure}[!htb]
  \centering
  \includegraphics[width=\textwidth]{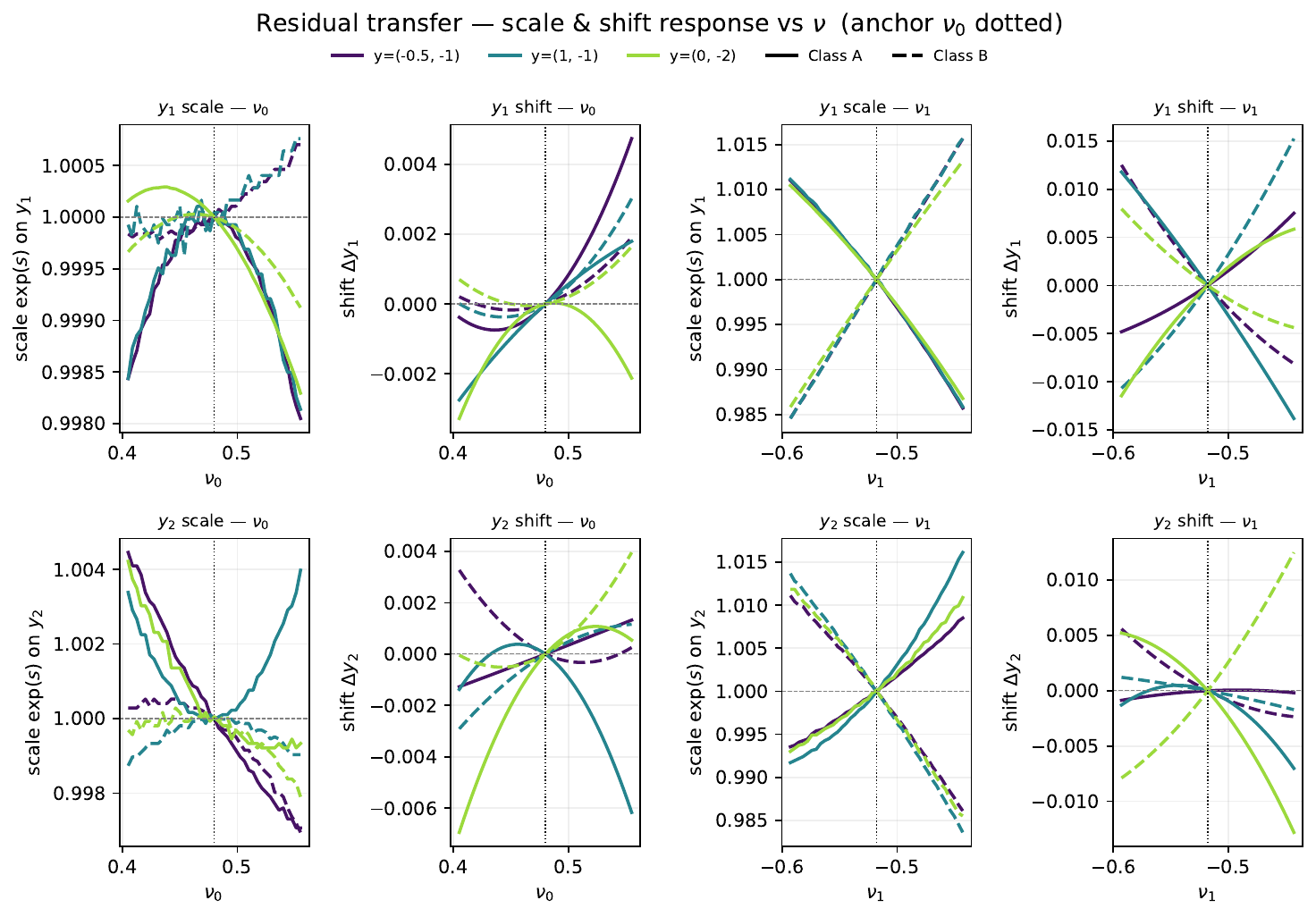}
  \caption{Response of the learned residual transformation $T_{\psi}$ as a function of the nuisances. Each panel shows the per-axis scale $e^{s}$ and shift $\Delta y$ that $T_{\psi}$ applies to the score components $y_1$ (top) and $y_2$ (bottom), as one nuisance is varied with the other held at the anchor, for three representative score points (colours) and the two classes (solid: A, dashed: B). The dotted line marks the anchor $\hat{\nu}$, where the residual reduces to the identity. The response is smooth and low-order in $\nu$, and opposite in sign for the two classes, as expected from the factorizable construction.}
  \label{fig:residual_response}
\end{figure}

Figure~\ref{fig:residual_response} shows the response of the trained residual transformation $T_{\psi}$: the per-axis scale $e^{s}$ and shift $\Delta y$ it applies to the score, as each nuisance is varied with the other held at the anchor. At the anchor $\hat{\nu}$ the residual reduces to the identity, with unit scale and zero shift, consistent with its construction as a deviation from the nominal map (Eq.~\ref{eq:Texp}). The response is smooth and low-order in $\nu$ over the training interval, which is what lets the amortized training of Section~\ref{sec:amortized} identify it from a coarse Gauss-Legendre grid, and it is opposite in sign for the two classes, as built into the dataset.

\section{Local covariance of the ensemble-averaged likelihood}
\label{sec:appendix_bma_hessian}

The systematic band and the orthogonal decomposition of Section~\ref{sec:results:step2} rely on the local Gaussian approximation of the ensemble-averaged likelihood at the best fit. We give here the form of its covariance, which makes explicit how the statistical uncertainty carried by the bootstrap ensemble enters the curvature.

Let $\ell_b(\nu) = -\ln L_b(\nu)$ be the profiled negative log-likelihood of member $b$, each rebased to its own optimum as in Section~\ref{sec:stat_uncertainty}, so that the ensemble-averaged objective of Eq.~\ref{eq:bma} reads
\begin{equation}
  S_{\mathrm{ens}}(\nu) = -\ln\!\left[\frac{1}{K}\sum_{b=1}^{K} e^{-\ell_b(\nu)}\right].
\end{equation}
With the per-member gradients $g_b = \nabla_\nu \ell_b$ and Hessians $H_b = \nabla_\nu^2 \ell_b$, and the likelihood weights
\begin{equation}
  w_b(\nu) = \frac{e^{-\ell_b(\nu)}}{\sum_{b'} e^{-\ell_{b'}(\nu)}}, \qquad \sum_b w_b = 1,
\end{equation}
differentiating $S_{\mathrm{ens}}$ twice gives the Hessian of the averaged objective,
\begin{equation}\label{eq:bma_hessian}
  H_{\mathrm{ens}} = \underbrace{\sum_b w_b H_b}_{H_{\mathrm{within}}} - \underbrace{\Big(\sum_b w_b\, g_b g_b^{T} - \bar{g}\,\bar{g}^{T}\Big)}_{\mathrm{Cov}_w[g]}, \qquad \bar{g} = \sum_b w_b g_b,
\end{equation}
with the nuisance covariance $C_{\mathrm{ens}} = H_{\mathrm{ens}}^{-1}$ evaluated at the best fit $\hat{\nu}$, where $\bar{g} = 0$. The gradients and Hessians are obtained by automatic differentiation.

The two terms separate the two sources of uncertainty. The first, $H_{\mathrm{within}}$, is the weighted-average curvature of the individual members: it is the Hessian one would obtain if the transformation were known exactly, and its inverse $C_{\mathrm{within}} = H_{\mathrm{within}}^{-1}$ is the systematic-only covariance. The second, the weighted covariance $\mathrm{Cov}_w[g]$ of the per-member gradients, is positive semi-definite and therefore reduces the curvature, inflating $C_{\mathrm{ens}}$ relative to $C_{\mathrm{within}}$. This inflation is the scatter of the member optima, the finite-sample uncertainty of the Step 1 transformation propagated to the nuisances; it is the local-Gaussian counterpart of the broadening produced by averaging the likelihoods in Eq.~\ref{eq:bma}. The two ellipses of Figure~\ref{fig:ortho_hessian} are the $C_{\mathrm{ens}}$ and $C_{\mathrm{within}}$ contours, and the eigenvectors visualized in Figure~\ref{fig:ortho_pull} are those of $C_{\mathrm{ens}}$.

\section{Cross-terms and the correlated systematic model}
\label{sec:appendix_crossterms}

The Factorizable Normalizing Flows of Ref.~\cite{Valsecchi:2026fnf} carry the systematic dependence of the input densities through affine scale and shift parameters that are polynomial in the nuisances and additive across them: each nuisance acts through its own coefficients, and the joint response is recovered at inference by summation, without sampling the combinatorially large joint space. This makes training scale linearly with the number of nuisances, and it mirrors the standard template-variation approach, in which each uncertainty is described by its own $\pm 1\sigma$ variations and correlations between sources are not modelled. The factorization is exact when the systematics act independently and deviates only through the neglected pairwise interactions, the \emph{cross-terms}: the optional bilinear $\nu_i\,\nu_j$ coefficients of Eq.~\ref{eq:fnf_poly}. For the two nuisances used here the single cross-term carries the $\nu_{\text{shift}}\,\nu_{\text{squeeze}}$ structure, vanishing on the axes, where only one nuisance is active, and growing toward the corners of the nuisance plane, where both act at once. Ref.~\cite{Valsecchi:2026fnf} quantifies this on the same dataset: measured by the per-event Kullback-Leibler divergence from the optimal likelihood, the factorized model closes on the optimum along the training axes and at the nominal point, the un-modelled interaction grows in the corners, and enabling the cross-terms, constrained by combined-variation samples, suppresses it across the whole plane.

This is the mechanism behind the coverage result of Section~\ref{sec:results:coverage}. The injected truth $(\nu_{\text{shift}}, \nu_{\text{squeeze}}) = (0.5, -0.5)$ sits where both nuisances are large and simultaneously active, the regime in which the factorized approximation is least accurate. The correlated deformation it omits is absorbed by the Step 1 fit into a small, compensating shift of the best-fit nuisances, biasing the central estimator while leaving the variance extracted in Step 2 correctly calibrated. Enabling the cross-terms in the kinematic and score input flows, calibrated on combined $\pm 1\sigma$ variations, restores the joint response and removes the bias, as the right panel of Figure~\ref{fig:coverage_test1} shows.

The same choice arises for the DoI residual $T_{\psi}$, built with the same factorizable construction so that its cross-terms can be enabled independently of those of the input model. Throughout this work, including the coverage run above, only the input model $T_{\chi}$ is augmented with cross-terms while $T_{\psi}$ is kept factorized: enabling them in $T_{\psi}$ would capture a second-order response of the measurement itself, distinct from the correlated data deformation the input cross-terms describe, and the coverage bias is already removed by the latter. The cost is modest: each correlated pair adds its own coefficients and combined-variation samples, so the count grows with the number of nuisance \emph{pairs}, a quadratic rather than exponential cost, and can be enabled selectively for the pairs that matter. We use the factorized model as the conservative default for the main results, invoking the correlated model in the coverage study to show that the residual bias is a modelling effect that precise systematic modelling removes.

\end{appendices}

\bibliographystyle{unsrtnat}
\bibliography{my_bibliography}

\end{document}